\documentclass[preprint,10pt]{elsarticle}

\biboptions{numbers,sort&compress}
\usepackage{amsmath} 

\usepackage{tabularx}
\usepackage{float}
\usepackage[normalem]{ulem}
\usepackage{orcidlink}

\usepackage{xurl}

\def\tsc#1{\csdef{#1}{\textsc{\lowercase{#1}}\xspace}}
\tsc{WGM}
\tsc{QE}
\tsc{EP}
\tsc{PMS}
\tsc{BEC}
\tsc{DE}

\author[1]{Pierrick Pochelu \orcidlink{0000-0002-3525-5033} \corref{cor1}\fnref{fn1}}
\ead{pierrick.pochelu@uni.lu}

\author[2]{Hyacinthe Cartiaux \orcidlink{0000-0002-4133-9065} \fnref{fn3}}
\ead{hyacinthe.cartiaux@uni.lu}

\author[2]{Julien Schleich \orcidlink{0000-0002-8406-1251} \fnref{fn3}}
\ead{julien.schleich@uni.lu}

\cortext[cor1]{Corresponding author}

\address[1]{LuxProvide S.A., Bertrange, Luxembourg}
\address[2]{University of Luxembourg, Esch-sur-Alzette, Luxembourg}

\begin{document}
\let\WriteBookmarks\relax
\def\floatpagepagefraction{1}
\def\textpagefraction{.001}

\title{What Artificial Intelligence can do for High-Performance Computing systems?} 

\begin{abstract}

High-performance computing (HPC) centers consume substantial power, incurring environmental and operational costs. This review assesses how artificial intelligence (AI), including machine learning (ML) and optimization, improves the efficiency of operational HPC systems. Approximately 1,800 publications from 2019 to 2025 were manually screened using predefined inclusion/exclusion criteria; 74 “AI for HPC” papers were retained and grouped into six application areas: performance estimation, performance optimization, scheduling, surrogate modeling, fault detection, and language-model-based automation. 

Scheduling is the most active area, spanning research-oriented reinforcement-learning schedulers to production-friendly hybrids that combine ML with heuristics. Supervised performance estimation is foundational for both scheduling and optimization. Graph neural networks and time-series models strengthen anomaly detection by capturing spatio-temporal dependencies in production telemetry. Domain-specialized language models for HPC can outperform general-purpose LLMs on targeted coding and automation tasks. Together, these findings highlight integration opportunities such as LLM-based operating-system concepts and underscore the need for advances in MLOps, standardization of AI components, and benchmarking methodology.

\end{abstract}


\begin{keyword}
Artificial Intelligence \sep High-Performance Computing \sep Software Performance
\end{keyword}

\maketitle

\section{Introduction}
High-performance computing (HPC) has become an essential tool for advancing scientific research and industrial applications, providing the computational power needed to solve complex problems in fields such as simulation, machine learning (ML), and big data processing. However, the increasing complexity of these systems due to growing heterogeneity and scale to meet rising demand presents significant challenges in optimizing job scheduling \cite{garlsched:2022, DRAS:2022, rlschedc:2021}. Traditional algorithms like First-Come-First-Serve \cite{sched:2022, slurm:2024}, Shortest Job Next, and Round-Robin often fail to efficiently allocate jobs to resources, leading to sub-optimal performance and wasting power consumed \cite{etp4hpc}. As HPC environments scale up, they also become more susceptible to hardware or software failures, disrupting ongoing tasks and necessitating re-scheduling or job restarts, further straining resources. 

{   This complexity is further compounded by the threat of cyber-attacks \cite{firewall:2024}, particularly in HPC systems exposed to public networks. Recent advances in LM-based tools show potential in supporting HPC cybersecurity engineers by automatically generating scripts and configuring firewall rules \cite{pracesec:2024}. However, cybersecurity within HPC remains a niche research domain, with limited empirical evidence on the deployment and effectiveness of such approaches in production environments.}

Additionally, the rapid pace of innovation in areas like accelerator architectures and large-scale AI (Artificial Intelligence) applications, such as Large Language Model (LLM) training \cite{deepspeed:2020}, inundates HPC teams with vast amounts of information \cite{litterworkflow:2020}, making automation increasingly necessary to support scientists and optimize system management.

{   Artificial Intelligence (AI) in this paper refers to the broad class of computational methods and algorithms that enable machines to perform tasks traditionally associated with human intelligence, such as learning, reasoning, decision-making, and language understanding. This definition encompasses machine learning (ML)—including supervised, unsupervised, reinforcement, and deep learning approaches. The definition is stretched to non-ML including optimization techniques such as evolutionary algorithms, Bayesian optimization, integer linear programming, and heuristics.

While machine learning currently dominates AI-for-HPC research, the definition of AI remains broader, encompassing both ML and non-ML approaches—from research prototypes to production systems. In practice, heuristic and rule-based algorithms often remain easier to deploy and maintain, as they avoid the complexities of building data pipelines, handling dataset shift, or tuning hyperparameters. In contrast, ML methods can deliver strong performance but require significant overhead for data curation, model tuning, and robustness monitoring.

Two sub-fields illustrate this contrast. Tensor parallelism optimization frameworks like Alpa \cite{alpa:2022} and Colossal-AI \cite{colossal:2023} automate decisions using Integer Linear Programming and empirical profiling, without relying on internal ML models. Conversely, Colossal-Auto \cite{syncolossalauto:2023} employs a machine learning estimator to approximate performance and reduce search costs. In scheduling, many research systems adopt hybrid approaches. For example, GARL \cite{garlsched:2022} combines reinforcement learning with expert-driven heuristics, while other schedulers blend data-driven estimators with constraint-based optimization \cite{sensetimechara:2021}.

}

This review is primarily aimed at entry-level scientists and engineers who are new to the challenges of current HPC systems. In this review, {  the contributions of ML and heuristic optimization are emphasized in addressing problems that are difficult to solve with conventional methods}, such as scheduling methods adapting to the dynamic nature of user activity with reinforcement learning  \cite{garlsched:2022, DRAS:2022, rlschedc:2021}, automating decision-making in environments with massive, unstructured datasets (code operations \cite{llm:2024, codex:2023, deepseek:2024}, graph dataset \cite{graafe:2024, anomalydetect:2023}, ... ). The application of ML in these areas serves as a complementary evolution of existing techniques, rather than a replacement.

As AI continues to evolve and mature, including ML algorithms and optimization methods, it presents promising opportunities to address the operational challenges inherent in HPC. The central research question guiding this review is: ``What can AI do for HPC systems?'' This question explores how AI techniques can be harnessed to support the HPC environment. The literature review aims to identify various areas for improvement.

\textbf{``What can AI do for HPC systems?''} This research question explores how AI methods can enhance the speed of applications, improve power efficiency, and reduce the operational costs of underlying HPC infrastructure. ML can distribute workloads more efficiently \cite{DAG:2024, rserv:2024, elasticflow:2023, panissara:2023, smdp:2023}  than traditional schedulers that rely on First-Come-First-Serve heuristics. Additionally, AI can substitute long computations by learning from input parameters and their corresponding outputs \cite{specinfer:2024, deepgp:2020, molecularsurog:2019}. Furthermore, it can monitor and automate infrastructure operations using statistical methods for anomaly detection \cite{diskfail:2024, graafe:2024, anomalydetect:2023, ruad:2022, anomalydetect:2022}, as well as streamline scripting and coding processes.

Given the proliferation of research in AI and HPC, it is crucial to map out and synthesize the vast body of knowledge to understand current trends and identify future directions. This review aims to provide a comprehensive overview of the latest advancements in AI techniques applied to HPC optimization and operations. Specifically, this article aims to achieve three main objectives:
\begin{itemize}
\item Comprehensive Review: Provides a detailed analysis of research papers across various AI applications in HPC, summarizing goals and key methods.
\item Evaluation of AI Techniques: Highlights the effectiveness of different AI methods, such as RL for scheduling, ML surrogates for simulation, and Language Model for coding, in improving HPC performance and efficiency.
\item Identification of Trends: Identifies emerging trends in hardware and AI methods and current challenges in AI for HPC.
\end{itemize}


Section~\ref{sec:background} reviews recent trends in HPC. {  Section~\ref{sec:method} provides details of the methodology and the classification of collected papers across sub-fields.} Section~\ref{sec:sum} summarizes key findings. Section~\ref{sec:integr} outlines a vision for potential AI integration in HPC. Finally, section~\ref{sec:con} concludes by synthesizing the current research directions and the integration challenges. 

\section{Background}
\label{sec:background}

\subsection{Definition of key terms}

AI refers to software capable of performing tasks requiring human intelligence, with methods including data-driven approaches, constraint satisfaction, and optimization algorithms \cite{turing:1950}. Recent advancements, particularly in deep learning, have broadened AI's scope across data applications supported by accelerators like GPUs and efficient tensor operations \cite{alexnet:2012}.

\subsection{HPC/AI convergence}

The convergence of HPC and AI can be framed through two main questions: ``What can HPC do for AI?'' and ``What can AI do for HPC?'' The first question focuses on how HPC can enhance AI by providing the computational power necessary to improve the accuracy of AI models, manage the memory requirements of large models with billions of parameters, and accelerate various stages of the AI lifecycle, including hyperparameter tuning, training, and inference. However, for the scope of this review, this question will be set aside to concentrate on the second question: ``What can AI do for HPC?'' This question explores how AI can be applied to optimize HPC infrastructure, including automatic parallelization and distribution of code, and how it can improve overall HPC system performance, efficiency, and sustainability \cite{etp4hpc}.

\begin{figure}[h!]
\centering
\includegraphics[width=\columnwidth]{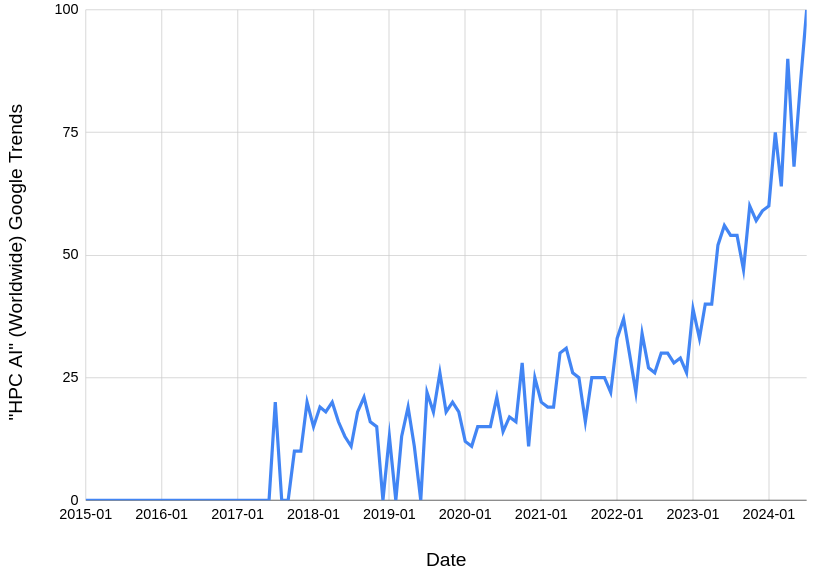}
\caption{World-wide Google queries with joint ``HPC'' and  ``AI'' terms according to Google Trends. The measurement is sampled every month. The vertical axis represents relative search interest, normalized on a scale from 0 to 100.}
\label{fig:gt}
\end{figure}

The recent popularity of the fields of HPC and AI is reflected in the approximately 20,000 scientific Google Scholar entries responding to the queries with joint ``HPC'' and ``AI'' terms.  Google Trends data in {   figure~\ref{fig:gt}} shows the dynamics of the interest in such requests.

The first measurable number of requests appeared in July 2017, and the frequency increased significantly at the end of 2022, coinciding with the release of ChatGPT LLM. {  The interest in the intersection of HPC and AI is not only increasing}, but the rate of increase itself is accelerating, driven by the rise of large AI applications. This growing interest suggests that the trend will continue to attract significant attention shortly. As more AI applications emerge that require high computational power, and as HPC systems are increasingly utilized, it becomes strategic to find smart ways to understand and manage these workloads. However, it is important to note that the general public may sometimes conflate single-node and multi-node HPC systems when querying about HPC/AI, which could lead to misunderstandings regarding the actual demand and interest in these infrastructures.

\subsection{Environmental Concerns and Opportunities}

\begin{figure}[h!]
\centering
\includegraphics[width=\columnwidth]{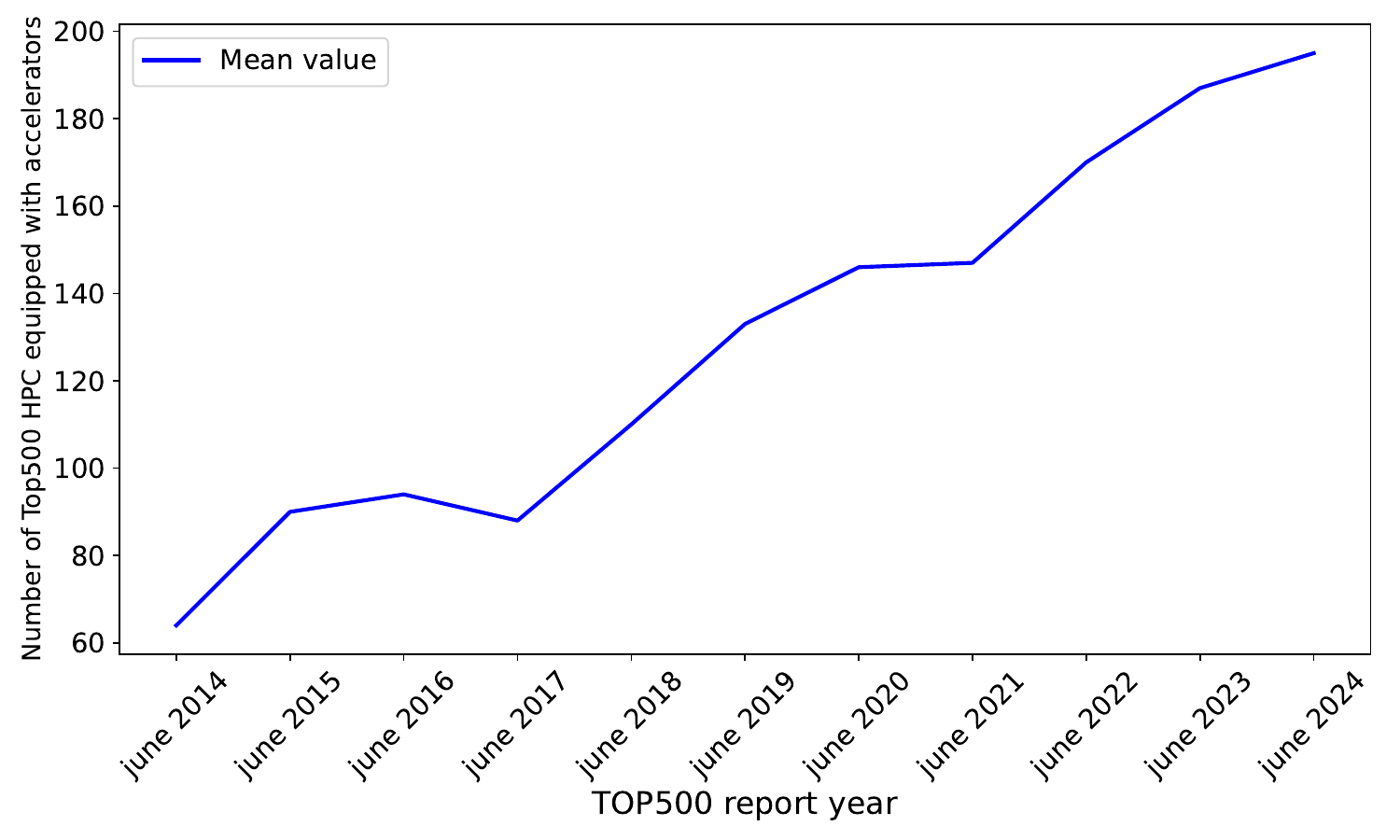}
\caption{Number of top 500 HPC systems \cite{top500} equipped with accelerators from June 2014 to June 2024. The number has increased from 64/500 in June 2014 to 195/500 in June 2024.}
\label{fig:b}
\end{figure}

\begin{figure}[h!]
\centering
\includegraphics[width=\columnwidth]{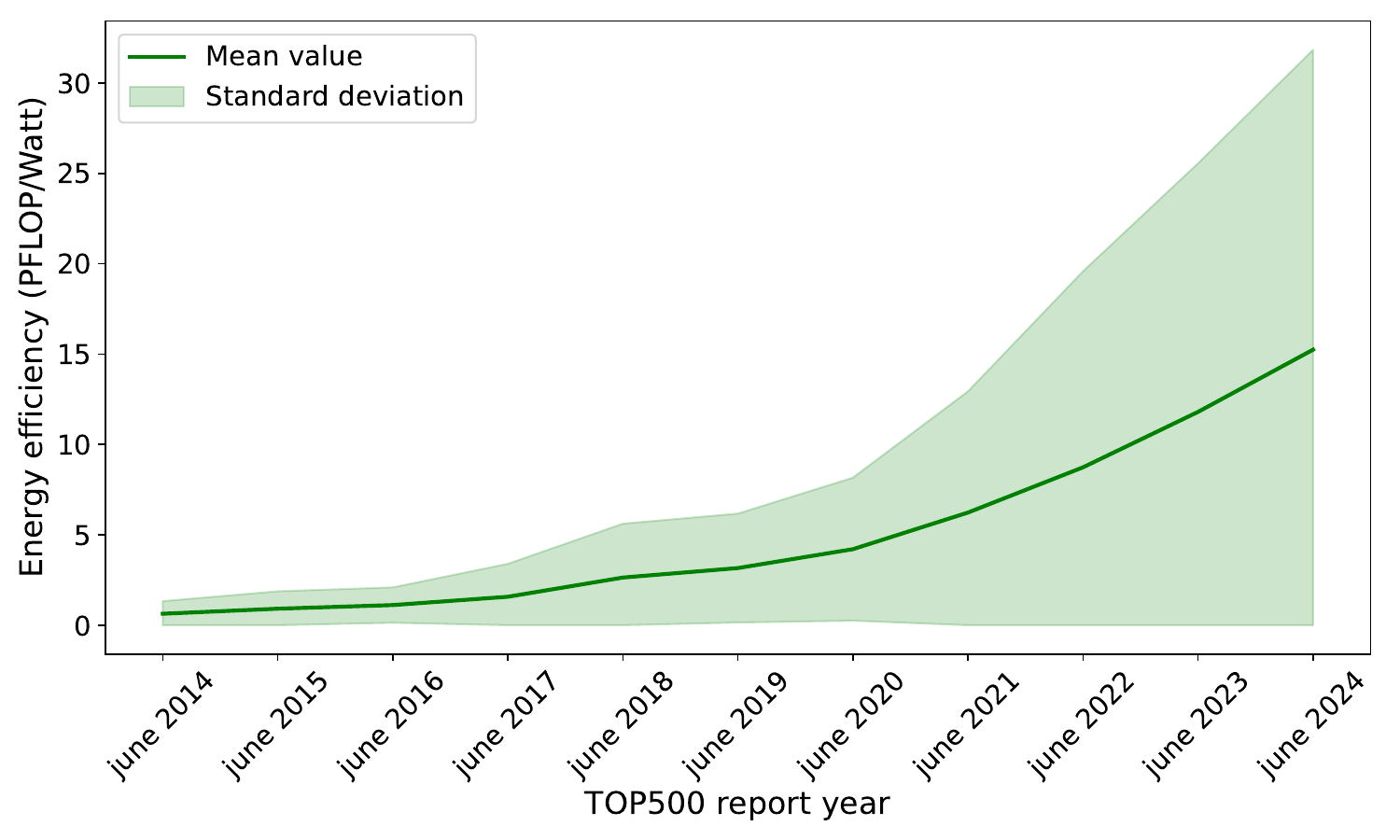}
\caption{Energy efficiency (PFLOPS/Watt) of top 500 HPC systems \cite{top500}  equipped with accelerators from June 2015 to June 2024. A higher average is better. The average has moved from 0.6 in June 2014 to 15.2 in June 2024.}
\label{fig:g}
\end{figure}

\begin{figure}[h!]
\centering
\includegraphics[width=\columnwidth]{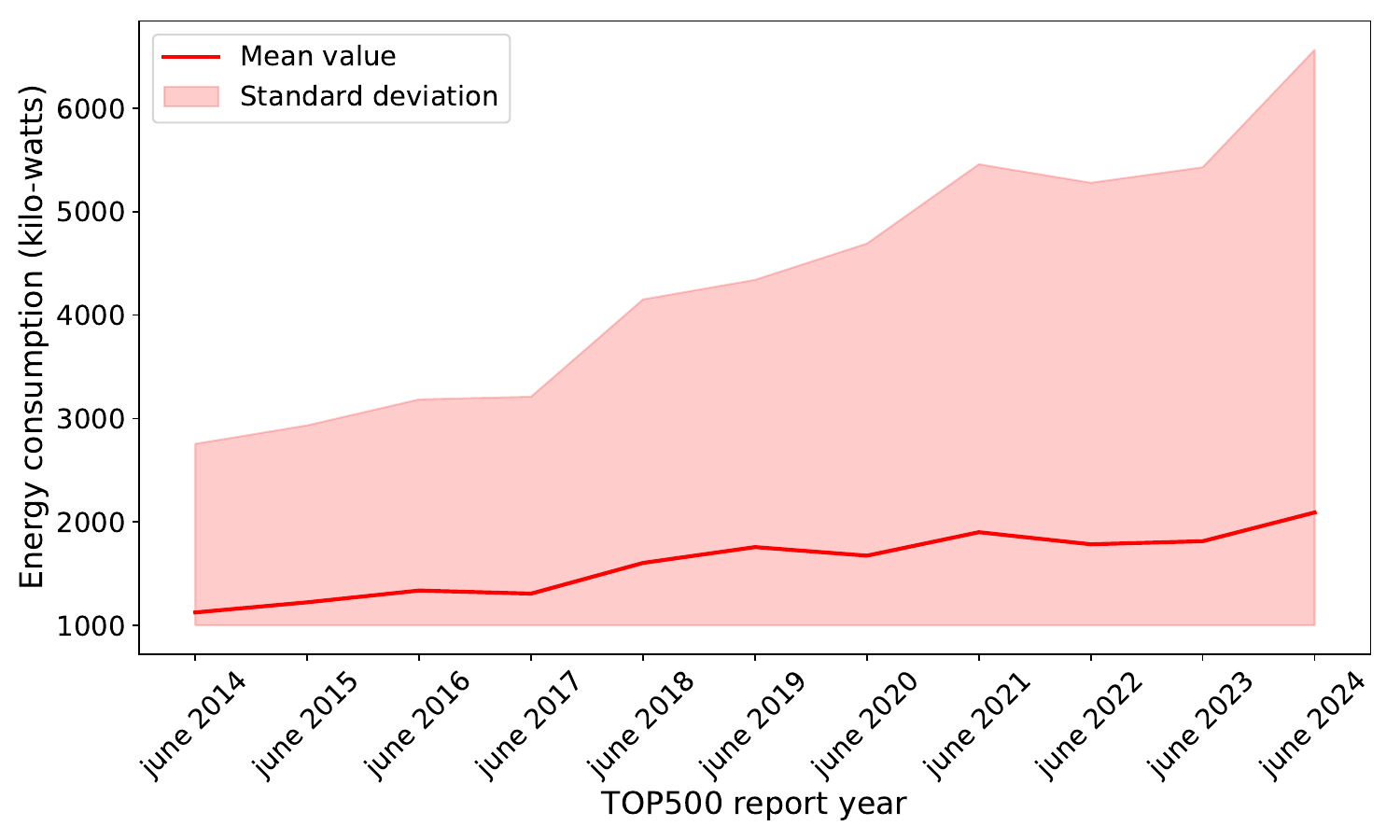}
\caption{Energy consumption (Kilowatts) of top 500 HPC systems \cite{top500} from June 2015 to June 2024. The average has increased from 1.1 MW in June 2014 to 2.1 MW in June 2024.}
\label{fig:r}
\end{figure}

Growing concerns around sustainability have led to increased emphasis on energy efficiency in HPC, as reflected in the Green-TOP500 list \cite{top500}. While the adoption of accelerators has improved performance per watt (\ref{fig:b}, \ref{fig:g}), overall energy consumption continues to rise due to increasing computational demands (\ref{fig:r}). To address this, AI-driven approaches such as intelligent scheduling, surrogate modeling, and real-time anomaly detection are being explored to enhance energy-aware resource management and system sustainability.

Despite the growing energy demands of HPC systems, their role in addressing environmental challenges is increasingly vital. For example, the Destination Earth initiative harnesses HPC to build a digital twin of the planet, supporting climate change adaptation and {  disaster risk} management through advanced Earth system modeling \cite{destin:2023}. In parallel, the U.S. Department of Energy’s HPC4EnergyInnovation program \cite{us:2024} connects HPC resources with industry to accelerate the development of clean energy technologies and materials.

{

\section{ What Can AI Do for HPC? A Review of Existing Surveys}
\label{sec:previous_surveys}

Recent surveys have explored the potential of integrating AI into HPC. Notably, some works~\cite{survworkflow:2024, survworkflow:2025} propose a conceptual taxonomy that distinguishes three modes of AI-HPC interaction: 
\begin{itemize}
    \item \textbf{AI-in-HPC}, where an AI model is used to replace a computationally intensive component, or the whole HPC simulation itself.
    \item \textbf{AI-out-HPC}, where AI is used to steer the HPC components or generate new data or parameterization.
    \item \textbf{AI-about-HPC}, where AI models are concurrent and coupled to the main HPC tasks and run synergistically with simulation.
\end{itemize}
This taxonomy emphasizes the physical or logical placement of AI within the HPC workflow rather than the specific use cases of AI models. While a general mapping between location and function is possible, ambiguity arises in scenarios in AI models addressing both application-level and system-level data sources and knobs \cite{bayes:2020, anomalydetect:2022}.

Another relevant perspective is provided by the "Learning Everywhere" paper~\cite{surveverywhere:2019}, which introduces another taxonomy:
\begin{itemize}
    \item \textbf{MLautotuning}, which involves automated tuning of runtime systems.
    \item \textbf{MLcontrol}, where ML guides simulation execution to optimize objective and data collection.
    \item \textbf{MLafterHPC}, referring to post-simulation analysis using machine learning.
    \item \textbf{MLaroundHPC}, which includes surrogate modeling.
\end{itemize}
This categorization is primarily focused on improving simulation workflows rather than the HPC system infrastructure. Additionally, it centers exclusively on machine learning, omitting other AI paradigms such as heuristics and Integer Linear Programming, which are still widely used in HPC research. Finally, the split between runtime system tuning (MLautotuning) and application-level simulation optimization (MLcontrol) complicates the classification of papers that simultaneously optimize across both layers, such as HiPerBOt \cite{bayes:2020}.

Other surveys concentrate on more specific aspects of AI in HPC. One survey~\cite{survscheduling:2023} explores energy-aware scheduling methods and emphasizes the increasing adoption of machine learning for multi-objective optimization. Another~\cite{survsoftware:2025} examines the influence of AI, particularly large language models, on HPC software development, highlighting both productivity improvements and concerns around trustworthiness. A separate study~\cite{survscalable:2025} presents scalable architectures for hybrid ML-HPC workflows, with a particular focus on runtime resource management and orchestration.

While these surveys identify promising directions, they also reveal a lack of integration across AI-HPC use cases. Brewer et al.~\cite{survworkflow:2025} explicitly call for benchmarks to evaluate AI-enhanced workflows, and Badia et al.~\cite{survworkflow:2024} underscore the absence of shared APIs and data formats that would allow workflow components to interoperate. This fragmentation makes it difficult to compose, reuse, or benchmark AI methods across different HPC subsystems. As noted by Skvortsov and Stupnikov~\cite{fairworkflow:2022}, adopting FAIR (Findable, Accessible, Interoperable, Reusable) principles in HPC workflows could address many of these issues, but practical implementation remains limited.

In contrast to prior work, our survey provides a more comprehensive overview of both software and operating HPC in daily activities. It includes HPC components for scheduling, performance estimation, fault detection, performance optimization, and scripting/code automation. By mapping interdependencies between AI applications and analyzing integration challenges, this survey aims to support a unified and intelligent HPC ecosystem.

}

\section{AI for HPC methods}
\label{sec:method}

The research question is given by \textbf{``What artificial intelligence can do for high-performance computing systems?''}, the following section discusses the methodology employed to research this question, and then, the {  capabilities} of AI are discussed.

\subsection{Research Methodology}

{  The research scope of this review is narrowed towards AI for HPC. To cover relevant literature, searches were conducted in engines such as IEEE Xplore, Scopus, Google Scholar, and Google.}

{  A comprehensive search strategy was employed using a combination of keywords with AND and OR logic to ensure comprehensive and relevant studies were captured.} 

The requests are the conjunction of 2 parts:
\begin{itemize}
    \item The AI-related method terms: ``artificial intelligence'' OR ``machine learning'' OR ``deep learning'' OR ``reinforcement learning'' OR ``LLM''.
    \item The HPC goal aimed: ``HPC'' OR ``software performance'' OR ``accelerator''.
\end{itemize}

{
 The following exclusion criteria were applied:
\begin{itemize}
    \item Papers published before 2018.  While valuable, these works were considered out of scope due to the rapid advancement of AI capabilities and the ongoing evolution of HPC architectures toward the exascale era.
    \item Papers lacking HPC- or AI-related keywords. Papers that did not mention these terms in their title or abstract were considered outside the scope of this review.
    \item ``HPC for AI'' papers, which are outside the scope compared to ``AI for HPC''. In some cases, reading the full article was necessary to determine the research question addressed.
    \item Papers on AI for micro-architecture design. This includes studies on chipset placement, logic synthesis, or quantum gate compilation. By contrast, such works fall within the scope of HPC if they address how AI improves the design of interconnected computing units or the synthesis of parallel computing architectures. This constraint allows the analysis to remain focused on operational aspects, rather than diverging into low-level physical design of computers.
    \item Only peer-reviewed documents were retained. Non–peer-reviewed sources may be cited for complementary discussion of trends or methodology, but were excluded from the core set of 74 reviewed papers.
\end{itemize}
}

{   In this study, approximately  1,800 papers were manually reviewed}, focusing on the intersection of AI and HPC.  Among these, { 823} were directly related to this intersection. 

{  Approximately 91\% of the papers} were centered on ``HPC for AI'' {  (around 749 papers), while only 9\% were focused on ``AI for HPC'' ({ 74} papers), with minimal overlap between the two categories due to aiming at different goals.} {  It is acknowledged that the specific focus of this review may be on a more limited subset of the HPC/AI area}, but the potential impact of this research is significant. ``HPC for AI'' generally improves the HPC utilization at the scale of a specific application set, and ``AI for HPC'' focuses on improving the overall HPC infrastructure to support a broader range of applications. The growing number of AI applications utilizing HPC, the exponential increase in demand for computational resources, and the environmental concerns related to HPC systems make research in this area increasingly urgent.

{  This review was conducted manually} after finding that machine learning-based approaches were insufficient for accurately categorizing scientific literature or extracting meaningful insights. Public LLMs, such as ChatGPT, often exhibit bias and produce out-of-context interpretations due to the predominance of ``HPC for AI'' publications over ``AI for HPC'' in the current literature. Additionally, machine learning methods that rely on word frequency and conventional search engines necessitate manual analysis to accurately classify documents, particularly in distinguishing between ``HPC for AI'' and ``AI for HPC''. {  Scientific impact was also prioritized by considering journal and conference metrics as well as document citations per year.}

{

Finally, to disambiguate papers spanning multiple subfields, an explicit tie-breaking rule was applied: classify by the last AI component in the end-to-end workflow (the “terminal-component rule”). For example, for a pipeline comprising (i) an AI-based performance estimator, (ii) an AI-driven performance optimizer, and (iii) an AI-enhanced scheduler, the paper is assigned to “AI scheduler.” This convention yields unique and reproducible assignments. The common synergies between subfields are the topic of section~\ref{sec:synergy}.

}

\subsection{Methodology Limitations}

{   Some limitations characterize the methodology. First, despite efforts to systematize the process, it is not fully automatic due to varying query capabilities and the maximum number of entries per page across digital libraries. Second, cross-validation was performed among co-authors to mitigate classification error or bias, but complete elimination of interpretative variance is not feasible. Third, the continuous evolution of the literature implies that a reproduction of the same search protocol at a later time may yield different results. This limitation is intrinsic to any snapshot-based survey but remains important to recognize in the context of a rapidly developing domain such as AI for HPC.

}

\subsection{Overview}

{  Among the papers that were collected addressing the research question, the aim was to identify appropriate groups to categorize the findings.}

To provide a more detailed and organized analysis, {  the `AI for HPC'' research was further divided into 6 categories:}
\begin{itemize}
    \item {  \textbf{AI for estimating the computing performance of a workload}. Papers in this category focus on leveraging AI to predict key performance metrics for jobs running on HPC systems, such as job duration, memory consumption, CPU usage, and overall resource demands. Accurate predictions of these metrics are crucial for efficient job scheduling, resource allocation, and system management. AI models can analyze historical data and job characteristics to provide detailed job profiles, which can then be used to anticipate how long a job will take to complete or how much memory it will require. These insights help optimize the scheduling of jobs, avoid resource contention, and improve the overall throughput of the system. Additionally, these predictions can inform other algorithms within the HPC environment, such as those used for load balancing, fault tolerance, or energy management.
    \item \textbf{AI for computational performance optimization of applications}: Papers in this category focus on using AI to enhance the computational performance of applications running on HPC systems. This involves optimizing various aspects of the application, such as computing speed, efficiency, and resource utilization. AI-driven techniques can automatically adjust application parameters, select the most suitable run-time libraries, and fine-tune compiler settings to maximize performance. By analyzing the specific needs of an application and the characteristics of the underlying hardware, AI can identify bottlenecks, predict optimal configurations, and adapt the execution environment dynamically. 
    \item \textbf{AI for Job/computer node Scheduling}:  Papers in this category focus on using AI to optimize the scheduling of jobs on computing resources. In HPC, a scheduler is responsible for managing the queue of jobs that need to be executed on the available compute nodes (the individual units within an HPC system that perform the computations). The scheduler's role is to allocate these jobs to the compute nodes in a way that maximizes resource utilization, reduces job wait times, and balances the workload across the system. AI techniques can enhance this process by predicting the most efficient allocation of resources, dynamically adjusting schedules based on real-time conditions, and minimizing idle time on the compute nodes. }
    \item \textbf{Machine Learning Surrogate Models for Long-Running Job Prediction}: Papers in this category investigate the use of machine learning surrogate models to predict the outcomes of long-running and computationally intensive jobs in HPC environments. These surrogate models act as simplified representations of the actual computational processes, allowing for faster computations without needing to run the full, resource-intensive HPC computations.
    \item \textbf{AI for Fault Detection}: This research area focuses on the application of AI techniques to identify and diagnose faults within the HPC infrastructure or distributed software systems. Faults in such environments can range from hardware malfunctions, such as failing compute nodes or storage devices, to software bugs or misconfigurations. AI models are trained to detect anomalies by analyzing vast amounts of operational data, including system logs, performance metrics, and sensor data, to identify patterns indicative of potential failures. Beyond just detecting faults as they occur, some studies advance the field by using AI to predict and anticipate faults before they happen, based on early warning signals, and some others attempt to provide a diagnosis of the fault. This proactive approach enables preemptive maintenance, therefore minimizing downtime and improving the overall reliability and resilience of the HPC system. 
    \item \textbf{{ Language Model (LM)} for Automation}: This area focuses on leveraging{ LM} techniques to automate scripting, coding, and other text-based tasks within HPC environments. LM-powered tools can assist various communities working in HPC centers. For instance, HPC administrators can use these tools to automatically generate scripts for system management, monitoring, and configuration tasks, reducing the manual effort required for routine operations. HPC researchers benefit from { LM}by using it to scrape, organize, and summarize multidisciplinary research articles, which helps them stay updated with the latest developments across various fields. Meanwhile, HPC scientists can utilize { LM}to streamline the development of complex codes, including those that employ distributed computing frameworks. By automating these tasks, { LM}enhances productivity, reduces errors, and allows HPC professionals to focus on more strategic and innovative aspects of their work.
\end{itemize}

\begin{figure}[h!]
\centering
\includegraphics[width=1.1\linewidth]{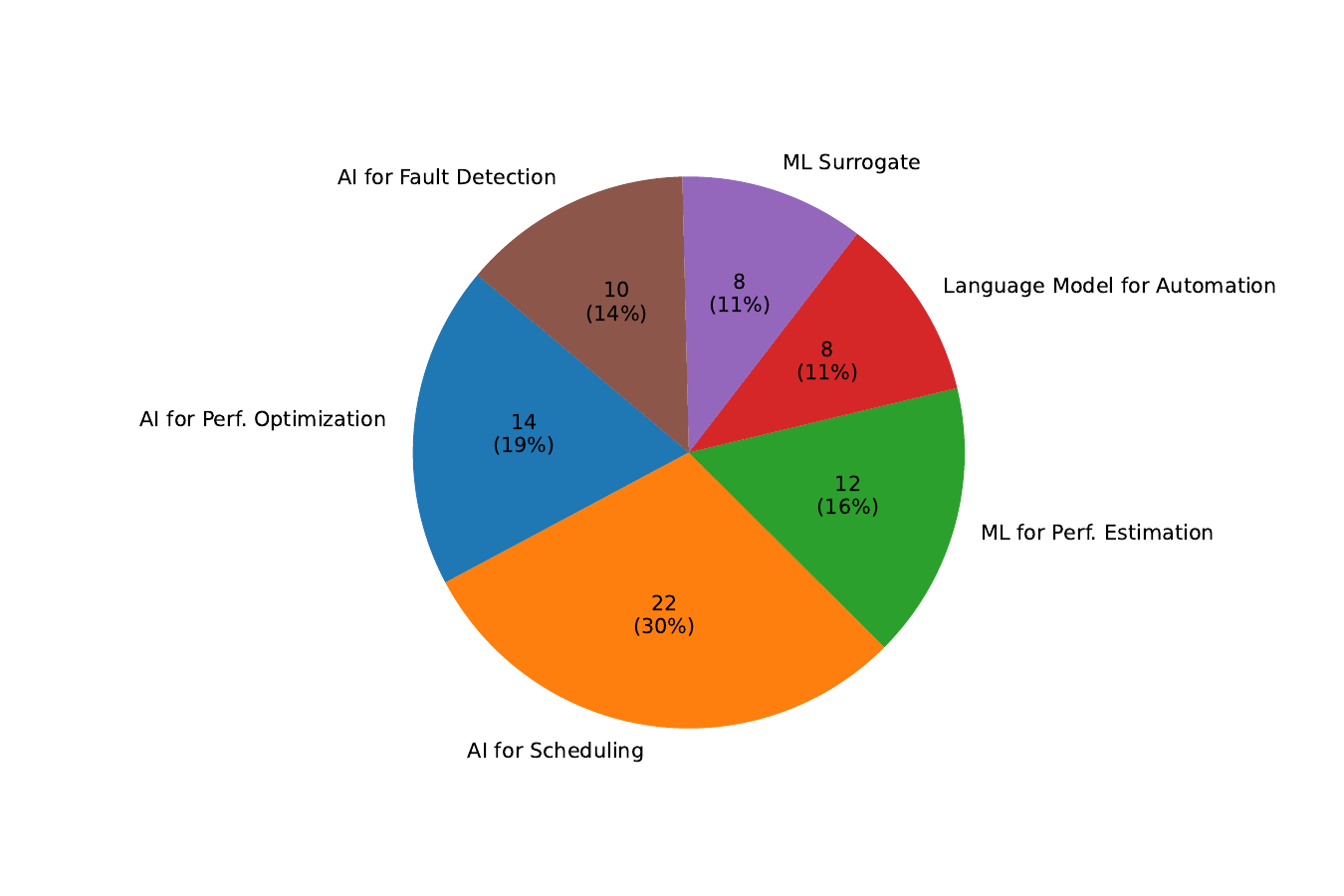}
\vspace{-2cm}
\caption{Distribution of {74} recent papers identified answering ``What AI can do for HPC systems?'' into 6 main categories}
\label{fig:distrib}
\end{figure}

Figure~\ref{fig:distrib} illustrates the distribution frequency of the {74} recent papers into 6 main categories addressing ``What can AI do for HPC?''. AI for performance estimation is grouped into a single category due to its extensive coverage and debatable split of some papers between infrastructure (HPC jobs computing time prediction) and the application-level view (specific application computing time prediction). 

To complement the quantitative overview of the research output and its impact, {  the following statistics have been gathered}. According to SJR Scopus data, of the { 74 }papers analyzed, 38\% are published in journals and conferences beyond the H-100 index, 40\% fall between the H-100 and H-10 range, 10\% are cited below H-10 but remain indexed, and 12\% are not indexed. The number of papers published between 2024-2023 appears to be relatively stagnant compared to those published during 2022-2020, indicating no clear trend among the six identified areas. {  However, the dataset of papers may be too limited to reliably identify or confirm such trends.}

In the next section, the articles are sorted by publication date to highlight the progression and recent advancements in the field.

\subsection{AI for Performance Estimation}

Computational performance encompasses various metrics, such as computing time and resource utilization. Accurate performance prediction before executing experiments in HPC is critical for avoiding the submission of inefficient or poorly performing workloads (e.g., anticipating resource exhausted error \cite{microsoft:2020}), inform a scheduler based on estimated completion times \cite{speccpu:2019}   (e.g., Shorter-Job-First, Shortest-Queue-Time heuristics), informing of the infrastructure load \cite{jobqueuepred:2023} \cite{uncertainty:2024}.  AI methods, particularly machine learning and deep learning, have demonstrated significant promise in accurately predicting these performance metrics.  The table~\ref{tab:esti} and \ref{tab:esti2} review several key studies that have applied AI techniques to performance estimation in HPC environments.

\begin{table}[H]
\scriptsize
\setlength{\tabcolsep}{0.05cm}
\caption{Studies on AI methods for performance prediction in HPC between 2019 to 2022}
\label{tab:esti}
\begin{tabularx}{\textwidth}{>{\centering\arraybackslash}p{0.04\textwidth}>{\centering\arraybackslash}p{0.16\textwidth}>{\centering\arraybackslash}p{0.2\textwidth}>{\raggedright\arraybackslash}p{0.2\textwidth}>{\raggedright\arraybackslash}p{0.4\textwidth}}
\textbf{Ref} & \textbf{HPC Goal} & \textbf{AI Method} & \textbf{Data source} & \textbf{Outcomes}  \\
\cite{speccpu:2019}                                 & Anticipating CPU utilization utilizing HPC Scheduler data                                                                                                                                & Ensemble of ML classifier  & Generated with LSF, 10,000 entries, features: user id, resource requirement: CPU cores, memory &  N/A                                                                                                                                                      \\
\cite{rnnhpc:2019}                                  & Classifying jobs for future scheduler                                                                                                                                                             & LSTM \cite{lstm} & synthetic jobs (with more than 500 runs)  & 100\% accuracy on short jobs, 96\% on long jobs, report challenge in tuning hyperparameters \\      
\cite{microsoft:2020}                                & GPU Memory consumption                                                                                                                                                                   & Deep Learning regressor & 5 models with different hyperparameters under 3 frameworks (TensorFlow, PyTorch, and MXNet) &  average errors of 11.8\%
(TensorFlow), 13.85\% (PyTorch), and 8.9\% (MXNet)   \\
\cite{sourcechara:2021}                              & Ensemble to predict DSP software performance based on code                                                                                                                               & Ensemble of ML regressor & the execution time of a set of functions selected from PHY DSP Benchmark and run them on TIC64 DSP processor &    average absolute relative prediction error is less than 10\%   \\
\end{tabularx}
\end{table}

\begin{table}[H]
\scriptsize
\setlength{\tabcolsep}{0.05cm}
\caption{Studies on AI methods for performance prediction in HPC between 2023 to 2025}
\label{tab:esti2}
\begin{tabularx}{\textwidth}{>{\centering\arraybackslash}p{0.04\textwidth}>{\centering\arraybackslash}p{0.16\textwidth}>{\centering\arraybackslash}p{0.2\textwidth}>{\raggedright\arraybackslash}p{0.2\textwidth}>{\raggedright\arraybackslash}p{0.4\textwidth}}
\textbf{Ref} & \textbf{HPC Goal} & \textbf{AI Method} & \textbf{Data source} & \textbf{Outcomes}  \\
\cite{icpe:2023}                                     & GPU code performance prediction                                                                                                                                                          & Bespoke Deep Learning regressor & SPEC 2017 benchmark  & R2 score 0.98                                                                                                                                         \\
\cite{jobqueuepred:2023}                             & Jobs Queuing Time Prediction                                                                                                                                                             & PCA-based Features Aggregation, Features Selection, classifier/regressor for time prediction, Bayesian Neural Network classifier, aleatoric/epistemic uncertainty & 34 days of data collected from PBS,  1 year and 8 months on the KIT FHII
System & Challenges: context shift, intrinsic data noise \\
\cite{aipaca:2024}                                   & Cloud Computing deep learning task cost                                                                                                                                                  & Deep Learning regressor & The tool contains random neural network sampling & N/A                                                                                                                                          \\
\cite{xaihpcops:2024}                                & XAI for HPC Operations                                                                                                                                                                   & HPC operating values, SHAP \cite{shap:2017} & historical dataset from onboard sensors: 12.5M samples, 26 features from energy and data. 310k samples, 9 features from the cooling system. 35.5k samples 22 features from environmental data &  global feature impact analysis, waterfall and force plots for local and global prediction, and dependency plots to capture feature interactions \\                                                                                                                                     \\
\cite{classiftime:2024}                              & ML classification to estimate jobs computing time, data drift estimation   & XGBoost \cite{xgboost:2016} classification, K-means  &  200 jobs executed using
Altair Radioss solver, 150 jobs executed
using the Altair Abaqus solver &  Integrated tool within the
Altair Access Portal, Frechet Inception Distance measurement  \\ 
\cite{uncertainty:2024}   & Quantifying Uncertainty in Job Queue Time Predictions   & XGBoost \cite{xgboost:2016} & 2.2 million jobs submitted to Eagle supercomputer operated by the National Renewable Energy Laboratory (NREL) in 2023 & Interval Conformance Rate of 16 cluster partitions is 0.646 \\
\cite{cmdpp:2024}        & Device-Model Agnostic Framework for Latency Prediction of Tensor Programs & Representation of tensor named as Compact Abstract Syntax Tree, Transformer \cite{transformer:2017}, KMeans-based sampling, Central Moment Discrepancy regularization & TenSet \cite{tenset:2021} + custom collection of 120 deep learning. Experiments on 5 GPU models & 14.03\% prediction error for cross-model. 10.85\% prediction error for cross-device prediction. Available open source. \\
\cite{multimodalestim:2025} & Multimodal task resource prediction & Informer \cite{informer:2021}, LSTM, GNN & 12 months production data sampled every 5min. Input: CPU util., mem. usage, storage, network, task exec. time, task queue length, node temperature, system-level failure, restart events & Achieved up to 89.9\% accuracy on runtime, CPU, memory, and storage predictions
\end{tabularx}
\end{table}

Performance estimation may play a pivotal role in making go/no-go decisions before executing computing tasks. It raises essential questions such as: is the workload worth the computing time, budget \cite{aipaca:2024}, and power consumption? Moreover, understanding whether a model can effectively fit the data, as discussed in \cite{microsoft:2020}, can introduce uncertainty, leading to time wasted in trial-and-error processes and delays caused by unnecessary memory allocation and crash report diagnostics.

Given the high dimensionality and nonlinearity involved in characterizing workloads, supervised ML and unsupervised ML methods can effectively leverage data from previously submitted jobs to make predictions. A notable challenge researchers face is predicting computing jobs with significantly varying durations, ranging from seconds to days. Although this is fundamentally a regression problem, some discretize predictions into ranges \cite{speccpu:2019} \cite{classiftime:2024}, and some \cite{jobqueuepred:2023} cascades both a classifier for range selection and regressors for finer-grain prediction.

Some authors tackle the complexity of performance estimation by focusing on uncertainty estimation \cite{jobqueuepred:2023} \cite{uncertainty:2024}. Understanding these uncertainties can lead to more robust scheduling and resource allocation strategies by providing decision-makers and algorithms with optimistic and pessimistic value ranges.

Finally, interpretable AI methods, such as those utilizing SHAP \cite{shap:2017} (Shapley Additive exPlanations), enhance the understanding of the contributions of input features to the model's predictions \cite{xaihpcops:2024}. This interpretability is crucial for refining performance estimation models and gaining insights into the factors influencing computational efficiency.

Work with performance estimation as the main objective accounts for 19\% of the papers collected in this study. However, AI performance estimation allows for extracting key performance characterization needed by AI Scheduler, AI anomaly detector, and AI task optimizer computational tasks.

\subsection{AI for Performance Optimization}

Performance optimization in HPC involves improving the efficiency and effectiveness of computational tasks. Research in AI for optimization is divided according to the goal: general-purpose software and deep learning applications. General-purpose software optimization includes tuning compilers, runtime environments, and application parameters. Deep learning applications require optimizing batch sizes, determining the optimal distribution of tensor computations across accelerators, and reducing communication overhead between accelerators.

Performance optimization is the goal of 19\% of the papers collected in this study. Table~\ref{tab:opt} and \ref{tab:opt2} review several key studies that have applied AI techniques to performance optimization in HPC environments.

\begin{table}[H]
\scriptsize
\setlength{\tabcolsep}{0.05cm}
\caption{Studies on AI methods for performance optimization in HPC between 2020 and 2022}
\label{tab:opt}
\begin{tabularx}{\textwidth}{>{\centering\arraybackslash}p{0.04\textwidth}>{\centering\arraybackslash}p{0.16\textwidth}>{\centering\arraybackslash}p{0.2\textwidth}>{\raggedright\arraybackslash}p{0.2\textwidth}>{\raggedright\arraybackslash}p{0.4\textwidth}}
\textbf{Ref} & \textbf{HPC Goal} & \textbf{AI Method} & \textbf{Data source} & \textbf{Outcomes}  \\
\cite{bayes:2020} & Tune compiler, run-time, application & Bayesian optimization, Transfer Learning & Performance code including Kripke, Hypre, LULESH, and OpenAtom & 50\% fewer samples for Kripke in comparison to the compared approach, 2x the number of good conﬁgurations for LULESH. \\
\cite{fleet:2020} & Optimization of ML ensembles training on HPC & Greedy algorithm optimizing computer time & 50 models derived from
DenseNets and 50 from ResNets & speed up the ensemble training by 1.12-1.92X over the default training method, and 1.23-1.97X over the state-of-the-art framework designed for homogeneous ensemble training \\
\cite{flextensor:2020} & Optimize tensor computation speed on Heterogeneous System  & heuristic, machine learning, custom {  scheduling} & 12 different kinds
of tensor computations with total hundreds of test cases & In average 1.83x performance speedup on NVIDIA V100 GPU compared to cuDNN; 1.72x performance speedup on Intel Xeon CPU compared to MKL-DNN for 2D convolution; 1.5x performance speedup on Xilinx VU9P FPGA compared to OpenCL baselines; 2.21x speedup on NVIDIA V100 GPU compared to the state-of-the-art. \\ 
\cite{ONES:2021} & Deep learning task scheduling with online batch size optimization & Online evolutionary scheduling algorithm & 64 GPUs on
TACC’s Longhorn supercomputers. AlexNet, ResNet, VGG,
GoogleNet, Inception, and BERT with different numbers of samples. From 3.6k to 20k samples. & job completion time reduced by up to 45.6\% compared to Deep RL \\
\cite{enssched:2021} & Optimization of an ensemble of electrolyte simulations design & Upper Confidence Bound, Proxy-ML & $10^{5}$ molecules (SMILES strings format) from the QM9 dataset & scales to 65,536 CPUs and accelerates the
discovery rate for high-performance molecules by a factor of 100 over unguided searches. Available open-source. \\
\cite{merlin:2022} & Ensemble of simulations for building surrogate models & Data collection tool for surrogate ML & N/A & to enqueue 40 million simulations in 100 s, with a 30-millisecond per-task overhead that is independent of ensemble size. Open source and in production. \\ 
\cite{pochelu:2022} & Optimization of ML ensembles inference on HPC & Greedy algorithm optimizing validation score and inference time & CIFAR100, ImageNet & Scalable number of GPUs to vary the inference throughput \\
\cite{alpa:2022} & Automatic deep learning parallelism & Dynamic Programming, Integer Linear Programming & N/A & Alpa generalizes to models with heterogeneous architectures. Alpa achieves a 3.5x speedup on 2 nodes and a 9.7x speedup on 4 nodes. Alpa also generalizes to models without manual strategies and achieves an 80\% linear scaling efficiency on Wide-ResNet with 4 nodes. Available Open-source and in production.
\end{tabularx}

\end{table}

\begin{table}[H]
\scriptsize
\setlength{\tabcolsep}{0.05cm}
\caption{Studies on AI methods for performance optimization in HPC between 2023 and 2025}
\label{tab:opt2}
\begin{tabularx}{\textwidth}{>{\centering\arraybackslash}p{0.04\textwidth}>{\centering\arraybackslash}p{0.16\textwidth}>{\centering\arraybackslash}p{0.2\textwidth}>{\raggedright\arraybackslash}p{0.2\textwidth}>{\raggedright\arraybackslash}p{0.4\textwidth}}
\textbf{Ref} & \textbf{HPC Goal} & \textbf{AI Method} & \textbf{Data source} & \textbf{Outcomes}  \\
\cite{multitasktuning:2023} & Multi-task HPC applications optimization & DeepGP \cite{dgp:2013}, Transfer Learning, EGO algorithm & { 2 applications: DGEQRF routine of the PLASMA library, the
DGEMM routine of the SLATE library} & {  better optima found on DGEQRF compared to OpenTuner, HpBandster, EGO-GP, LCM} 
 \\
\cite{colossal:2023} & Automatic deep learning parallelism & Greedy Algorithm & GPT-2 model with 10 billion parameters on the Wikipedia dataset & 2.76x training speedup
on large-scale models. Open source and in production. \\
\cite{syncolossalauto:2023} & Deep-learning intra-operator parallelism; activation checkpointing & Symbolic profiling with Torch.fx \cite{torchfx:2022}; Cluster Detector (models bandwidth/latency/topology); intra-op parallelism ILP solver adapted from Alpa \cite{alpa:2022}; activation-checkpointing solver \cite{rotor:2024} & N/A & GPT-2 14.5B: 0.824 PFLOPS vs. 0.728 (Megatron-LM) and 0.715 (3D tensor parallelism) \\
\cite{autotuning:2024} & Auto-tuning CUDA kernels on a set of modern GPUs & Methodology for GPU auto-tuning design & N/A & Challenge: It uses random search as a baseline, the search
spaces must be well-formed \\ 
\cite{exegpt:2024} & Tensor optimization & Branch-and-bound & N/A & Compared to FasterTransformer, ExeGPT
achieves up to 15.2x improvements in throughput and 6x improvements in latency. Overall, ExeGPT achieves an average
throughput gain of 2.9× across twenty evaluation scenarios. \\
\cite{gmorph:2024} & Inference of multiple DNN & Multi-task learning addresses this problem by designing a multi-task model that shares parameters across tasks
based on a single backbone DNN & N/A & inference latency of multi-DNNs reduced by 1.1-3x. Available open-source.
\end{tabularx}

\end{table}

The collected research addresses three complex optimization problems: multi-objective computational performance and application-specific score functions (e.g., \cite{bayes:2020, pochelu:2022}) where AI generates objectives to follow (e.g., validation score, uncertainty measurements). Optimization of ensembles of medium-sized tasks (e.g., \cite{enssched:2021, merlin:2022, fleet:2020, ONES:2021, pochelu:2022}) or large models distributed over multiple accelerators (\cite{alpa:2022, colossal:2023}), where AI allows tuning the distribution of computations, accelerator micro-code performance optimization (e.g., \cite{bayes:2020, softperf:2020, autotuning:2024}).

\subsection{AI for Scheduling}

HPC scheduling involves the allocation and management of computational tasks on HPC resources to maximize the utilization of computational resources and minimize wait times. This includes deciding the order and timing of jobs to optimize resource utilization and performance. AI methods, particularly machine learning and reinforcement learning, have been increasingly applied to improve HPC scheduling, addressing challenges such as large input/output sizes, dynamic states, and the need for accurate predictions.

AI for scheduling constitutes 30\% of the papers collected in this study. Tables \ref{tab:sched}, \ref{tab:sched2}, \ref{tab:sched3} review several key studies that have applied AI techniques to scheduling in HPC environments.

\begin{table}[H]
\scriptsize
\setlength{\tabcolsep}{0.05cm}
\caption{Studies on AI methods for scheduling in HPC between 2020 and 2021}
\label{tab:sched}
\begin{tabularx}{\textwidth}{>{\centering\arraybackslash}p{0.04\textwidth}>{\centering\arraybackslash}p{0.16\textwidth}>{\centering\arraybackslash}p{0.2\textwidth}>{\raggedright\arraybackslash}p{0.2\textwidth}>{\raggedright\arraybackslash}p{0.4\textwidth}}

\textbf{Ref} & \textbf{HPC Goal} & \textbf{AI Method} & \textbf{Data source} & \textbf{Outcomes}  \\
\hline
\cite{ear:2020} & power accounting, energy control/optimization & linear model \cite{linearear:2011} & 8 applications, 26 nodes at Leibniz Supercomputing. Input: iteration time, power, memory transition per sec., cycles per sec. & From +2\% to +12\% Gflops/watt. Tested on 6480 nodes. Open source and SLURM plugin \\
\cite{onlineknn:2021} & General improving the scheduling system, Auto-correction of the user-requested time & Online Learning, KNN, Linear Regression & Custom HPC simulator HPCsim implemented based on real system workload inputs. SimGrid \cite{simgrid:2014}. &  can be used to enhance primary prioritizing and backfilling methods without being restricted by specific scheduling methods \\ 
\cite{rlsched:2021} & HPC Jobs/Resource Scheduling & RL framework for HPC (A2C, SAC, PPO)  & CQSim simulator \cite{cqsim:2013}  &  DRAS-CQSim encapsulates simulation environments, agents, hyperparameter tuning options, and different reinforcement learning algorithms. Open Source. \\ 
\cite{sensetimechara:2021} & Analysis of job traces and time prediction & ML Duration Prediction, Quasi-Shortest-Service-First  (QSSF) & real-world job traces from SenseTime. Publicly available. &  QSSF service to improve the average JCT by up to 6.5×. Cluster Energy Saving service, which improves overall cluster utilization by up to 13\%. \\ 
\cite{rlschedb:2021} & HPC Jobs/Resource Scheduling & Hierarchical Neural Network, Policy Gradient, Deep Q-Learning  & Real jobs : Theta at ALCF \cite{alcf:2024} 122k (max 1 day length), Cori at NERSC \cite{nersc:2024} 2.6M (max 7 days). Train on real and synthetic jobs. & outperforming
existing scheduling policies by up to 45\%  \\
\cite{rlschedc:2021} & HPC Jobs/Resource Scheduling & Actor-Critic, Kernel-based Neural Network  & Publicly available Parallel Workloads Archive  \cite{swfarchive:2014}: SDSC-
SP2, HPC2N, PIK-IPLEX, ANL Intrepid, Lublin-1 \cite{lublin:2003}, Lublin-2 &  learned models perform stably even when applied to unseen workloads, making them practical for production use. Open Source \\ 
\cite{ONES:2021} & Deep learning task scheduling with online batch size optimization & Online Evolutionary Scheduling Algorithm  & AlexNet, ResNet, VGG16, GoogleNet, Inception V3, pre-trained BERT  &  ONES can reduce the average JCT by up to 45.6\%, compared to state-of-the-art (DRL) methods. In addition to that, it allows preemption and elastic batch size. Open source. \\ 
\cite{rlschert:2021} & RL for Jobs Scheduling & Proximal Policy Optimization, RNN-based Remaining Time Predictor, Kill Policy  & The Vienna Ab Initio Simulation Package \cite{vienna:1997} & Average Slowdown Time is lower 8.41 compared to state-of-the-art (DeepRM \cite{deeprm:2016}) 9.83. The wait time for long jobs using RL is longer than DeepRM, and the wait time for short jobs is shorter, indicating that short jobs are more likely to be backfilled. Open Source 
\end{tabularx}

\end{table}

\begin{table}[H]
\scriptsize
\setlength{\tabcolsep}{0.05cm}
\caption{Studies on AI methods for scheduling in HPC between 2022 and 2023}
\label{tab:sched2}
\begin{tabularx}{\textwidth}{>{\centering\arraybackslash}p{0.04\textwidth}>{\centering\arraybackslash}p{0.16\textwidth}>{\centering\arraybackslash}p{0.2\textwidth}>{\raggedright\arraybackslash}p{0.2\textwidth}>{\raggedright\arraybackslash}p{0.4\textwidth}}
\textbf{Ref} & \textbf{HPC Goal} & \textbf{AI Method} & \textbf{Data source} & \textbf{Outcomes}  \\
\cite{DRAS:2022} & RL for Jobs Scheduling & Hierarchical Neural Network  & Theta at ALCF \cite{alcf:2024} and Cori at NERSC \cite{nersc:2024} & Outperforms the existing heuristic and optimization approaches by up to 50\%. Open Source. \\
\cite{enhancegreedy:2022} & Online scheduling of task batches on heterogeneous servers & Greedy Algorithms, RL   &  Google cluster trace and real application environment of 3,000 face detection tasks submitted randomly in batches using from 20 to 100 cloud servers. &  The algorithm improves system gain (measured as the value of completed tasks minus system operation costs) by approximately 10\% to 30\% compared to state-of-the-art methods (optRA, OSSA, HAMO, DRL). Provides a worst-case performance guarantee. \\
\cite{garlsched:2022} & Large-scale scheduling & Generative Adversarial Deep Reinforcement Learning, PPO-Clip, Rule-based Scheduling  & Lublin-256 \cite{lublin:2003} (10k tasks), SWF archive \cite{swfarchive:2014}: HPC2N (527k tasks), SDSC-BLUE (250k tasks), SDSC-SP2 (73.5k tasks) & learned models can perform stably even when applied to invisible workloads \\ 
\cite{liquid:2022} & GPU resource management platform for DL jobs & regression model-based method for job resource requirement estimation to avoid users over-allocating computing resources. & CNN for MNIST dataset, NeuMF \cite{neurmf:2017} with MovieLens 20M dataset, ResNet-50, Inception V3, VGG-16. & accelerate the job execution speed by 18\% on average and shorten the average job completion time (JCT) by 21\% compared with the existing solution (Kubernetes, Kubeflow, and HiveD)  \\
\cite{smdp:2023} & Online Scheduling, Large-action space, varying sizes without retraining & Maskable PPO \cite{mask:2020}  & Smdp branch of the
sched-rl-gym package, with the CompactRM-v0 OpenAI Gym
environment &  outperforms learning models from the literature and classic heuristics. Robust to changes in workload and cluster sizes, showing transfer works with changes of cluster size of up to 10×. Open source. \\
\cite{panissara:2023} & Deep Learning training Jobs/Resource placement & Similarity-Based Policy  & 21 common CNN on ImageNet, 1680 jobs. Features: workload characteristics (number of layers, parameters, FLOPS, ...), number of GPUs   & Reducing makespan  \\ 
\cite{eelas:2023} & ML Inference scheduling in HPC-Edge for Latency/Battery optimization & Greedy Algorithm driven by MIPS/Watt  & NITOS testbed part of the SLICES-RI &  overall energy efficiency of over 41.8\% compared to the off-the-shelf K8s scheduler \\ 
\cite{lucid:2023} & Interpretable deep learning workload scheduler & Primo \cite{primo:2022}, Decision Tree, GA\^{2}M \cite{GA2M:2013} & 3 production-level traces collected: 101K jobs with average duration 13k sec, 23.9k jobs with avg. duration 5.4 sec, and 12.4k jobs with avg duration 25.5sec. The first two are SenseTime trace and the third is Microsoft trace. & reduces the average job completion time by up to 1.3x compared with state-of-the-art pre-emptive scheduler Tiresias. Open source. \\ 
\cite{elasticflow:2023} & elastic resource management for distributed deep learning & greedy algorithm, buddy allocation & 6 DL tasks: ResNet50, VGG16, InceptionV3, BERT, GPT-2, Deep Speech 2 & Evaluation results on a 128-GPU
cluster show that ElasticFlow can increase the
number of jobs that can meet their deadlines by 1.46–7.65x
compared to existing state-of-the-art solutions. 
Integrated with PyTorch. Open Source. \\
\cite{oiko:2023} & Resource-recommendation system for
HPC applications in the cloud & Multi Layer Perceptron & 8 Amazon EC2 instance types, 2 have a GPU available. 8 applications: 5 from ARCHER \cite{archer:2019}, 3 Tensorflow. & The optimal instance type was chosen in 90\%
of the cases for seven out of eight applications, scoring a Mean
Absolute Percentage Error (MAPE) consistently below 20\%. \\
\end{tabularx}
\end{table}

\begin{table}[H]
\scriptsize
\setlength{\tabcolsep}{0.05cm}
\caption{Studies on AI methods for scheduling in HPC in 2024}
\label{tab:sched3}
\begin{tabularx}{\textwidth}{>{\centering\arraybackslash}p{0.04\textwidth}>{\centering\arraybackslash}p{0.16\textwidth}>{\centering\arraybackslash}p{0.2\textwidth}>{\raggedright\arraybackslash}p{0.2\textwidth}>{\raggedright\arraybackslash}p{0.4\textwidth}}
\textbf{Ref} & \textbf{HPC Goal} & \textbf{AI Method} & \textbf{Data source} & \textbf{Outcomes}  \\
\cite{kubernsched:2024} & Kubernetes jobs placement & ML-Based Workload Placement  & 3-tier web application scenario & integration of various components: PDUs to track
energy usage, remote management interfaces to track servers’ health state, hardware testing score \\
\cite{rserv:2024} & Reserve in advance resources for optimal job execution, based on duration, dependencies, and machine characteristics & Mix of Unsupervised and Supervised ML for predicting  & 3 independent probability distributions to emulate the job submissions distribution: exponential for arrival, multinoulli for the number of nodes, gaussian mixture for wall time & Optimized work-flow has a makespan of 16 minutes, compared to 46 minutes
of the non-optimized one. Integration to SLURM  \\ 
\cite{DAG:2024} & Graph of jobs placement & Actor-Critic RL, heuristics selection among several: First-Come-First-Serve, Shortest Processing Time, and Shortest Queue Time, ...  &  Randomly generated DAG. Scenario 1: The jobs (100 or 300) were generated using a normal distribution for
defining the arrival time of each composing task, Scenario 2: Each DAG
is composed of up to 10 tasks, and the DAGs are generated with an interconnection
probability of 0.2 or 1.0 between DAG layers & dynamically adapted the policies selection for different workload contexts, eventually outperforming the counterparts schedulers (FCFS, SQF, SPF, SAF, F1, F2, F3, F4, AC) \\ 
{  \cite{herasched:2025}} & Intelligent job scheduling and node allocation & Maskable PPO \cite{mask:2020}, Hierarchical RL & data from HPCSim (open source) &  outperformed 27 scheduling strategies; Reduced job wait time. Available open source. \\
\end{tabularx}
\end{table}

Out of the collected papers in this category, half (\cite{rlsched:2021, rlschedb:2021, rlschedc:2021, rlschert:2021, DRAS:2022, enhancegreedy:2022, garlsched:2022, smdp:2023, DAG:2024}) highlight the potential of Reinforcement Learning (RL) to transform automatic scheduling by outperforming traditional heuristics such as First-Come-First-Serve (FCFS). RL algorithms can learn from historical patterns, including the arrival times of previous jobs, job characteristics (such as CPU cores required, memory requests, and reservation times), and the current state of the infrastructure (including available memory and CPU cores). By leveraging this information, RL can adaptively optimize scheduling decisions, leading to more efficient resource utilization and improved performance in HPC environments.
 
However, the challenges in applying AI to HPC scheduling are multifaceted. Large input and output sizes and the dynamic nature of input/output characteristics can hinder the learning process. RL techniques face significant hurdles in implementation due to the large-scale nature of input/output. The modeling of the dynamic learning should support the dynamic dimensionality of the action space and state (e.g., adding/removing a server should not lead to retraining a new agent). Those systems based on RL often rely on the need for precise time predictions from other ML estimators and an accurate simulator to be trained before being deployed in real infrastructure.

Simpler scheduling techniques informed by machine learning have shown promise and are easier to integrate into HPC schedulers such as SLURM, PBS, and LSF. Many employ greedy algorithms enhanced with ML predictions for job characterization  \cite{sensetimechara:2021, onlineknn:2021, oiko:2023} or comparisons between similar jobs and hardware to infer computation times \cite{panissara:2023}. Additionally, some research focuses on energy-driven scores to schedule workloads, as seen in \cite{eelas:2023}.

Finally, interpretability in AI  \cite{xaihpcops:2024} using algorithms like SHAP allows for a better understanding of the contribution of input values to the scheduling decisions, which is critical for trust and transparency in AI-driven scheduling systems.

\subsection{ML Surrogate}

ML surrogates consist of machine learning models that approximate the behavior of more complex and computationally expensive simulations. These surrogates have emerged as powerful tools for avoiding unnecessary computation. They are especially useful within the HPC domain, where long and expensive computations can be avoided with a supervised learning approach.

Table~\ref{tab:sur} summarizes works found in this field.

\begin{table}[H]
\scriptsize
\setlength{\tabcolsep}{0.05cm}
\caption{Studies on Machine Learning Surrogate Model}
\label{tab:sur}
\begin{tabularx}{\textwidth}{>{\centering\arraybackslash}p{0.04\textwidth}>{\centering\arraybackslash}p{0.16\textwidth}>{\centering\arraybackslash}p{0.2\textwidth}>{\raggedright\arraybackslash}p{0.2\textwidth}>{\raggedright\arraybackslash}p{0.4\textwidth}}
\textbf{Ref} & \textbf{HPC Goal} & \textbf{AI Method} & \textbf{Data source} & \textbf{Outcomes}  \\
\cite{molecularsurog:2019} & ML surrogate of MPI/OpenMP Simulation & Neural Network Regression & the training dataset consisted of 4804 simulation conﬁgurations, with 2060 input parameters & ANN ANN-based regression model predicted Contact density (MSE=0.0000718), midpoint density (MSE=0.0002293), and peak density (MSE=0.0002306). Simulation time (around 12 h) and ANN surrogate model (around 0.25 s). Open source.\\
\cite{deepgp:2020} & ML surrogate of high-fidelity multi-physics simulation & DeepGP \cite{dgp:2013}, variational inference, variance decomposition with Sobol indices, parameter screening with Morris method & 4 custom datasets: input dimension from 9 to 91, training samples from 120 to 200 & Small MAE and MSE. DeepGP also accurately captures the uncertainty (about 0.6\%). DeepGP successfully explains 95\% or more of the variance in all 12 outputs. \\
{\color{black}\cite{sci:2023}} & {\color{black}Replace code region with a faster ML model} & {\color{black}Neural Network} & {\color{black}LLVM instructions to generate a tree-based data dependency graph. Evaluated on 11 functions: 3 from numerical solvers, 5 from PARSEC \cite{parsec:2011}, 3 from Exascale Computing Project \cite{ecp:2017}.} & {\color{black}There is 1.89×–16.8× speedup with a harmonic mean of 5.50× across all three types of applications, compared with the application performance on CPU (using all 40 cores).} \\
{\color{black}\cite{hydro:2023}} & {\color{black}Optimize tuning workloads in both the job-level and cluster-level granularities} & {\color{black}Model Shrinker, Maximal Update Parameterization \cite{mu:2021}} & {\color{black}Evaluated on 6 models: GPT-3XL, Transformer, MobileNetV3 Large, VGG-11, ResNet-18} & {\color{black}It reduces tuning makespan by up to 78.5× compared with Ray Tune and no reduction in tuning quality.} \\
{\color{black}\cite{cfd:2024}} & {\color{black}Library for large-scale training of AI models for Computational Fluid Dynamics} & {\color{black}Convolution auto-encoder \cite{cae:2016}, Convolution Defiltering Model \cite{cdm:2023} based on Diffusion Probability Model \cite{dpm:2015} for Super-Resolution} & {\color{black}Turbulent Boundary Layer \cite{tbl:2019} simulation dataset (8.3 TB)} & {\color{black}Training scaling efficiency of 96\% with 3,664 GPUs compared to the performance on 1 node. Mean drag prediction error of $\approx$5\%. With the diffusion model: $\approx$4\%.} \\
{\color{black}\cite{dynn:2024}} & {\color{black}Training more efficiently dynamic neural networks (DyNN) \cite{dynsurv:2022} by predicting access order of tensors} & {\color{black}Three parallel Multilayer Perceptrons and an output layer} & {\color{black}Training on 24,000 samples from six diverse models (Transformer, CNN, LSTM). Tested on 2,000 samples from 8 models.} & {\color{black}The proposed model outperforms unified virtual memory (UVM) and dynamic tensor rematerialization (DTR) by 3× and 2.1× respectively in terms of maximum batch size.} \\
\cite{specinfer:2024} & Faster and smaller LLM to mimic larger LLM & Speculative, Tree-based predicted tokens, verification in parallel & Prompts/questions from: Chatbot Instruction Prompts \cite{chatbotinstructionprompts:2023}, ChatGPT Prompts \cite{chatgptprompts:2023}, WebQA \cite{webqa:2022}, Alpaca \cite{alpaca:2023}, and PIQA \cite{piqa:2020} & SpecInfer outperforms existing LLM serving systems by 1.5–2.8× for distributed inference and 2.6–3.5× for offloading-based inference, while preserving generative performance. Open source. \\
{  \cite{raja:2025}} & Performance modeling & Polynomial Regression, Least Squares Regression & RAJAPerf benchmark data \cite{rajasuite:2024}  & 54\% better than kNN with 33\% less data
\end{tabularx}
\end{table}

Surrogate models have been extensively used as replacements for physics-based computing, such as in molecular dynamics \cite{molecularsurog:2019, deepgp:2020} and computational fluid dynamics \cite{cfd:2024}. They are also increasingly adopted to accelerate deep learning workloads, including hyperparameter tuning \cite{hydro:2023}, training \cite{dynn:2024, hydro:2023}, and inference \cite{specinfer:2024}. Furthermore, surrogates serve as efficient substitutes for computationally expensive code patterns in scientific applications \cite{sci:2023}.

Recent AI techniques have significantly advanced the development of surrogate models. Convolutional autoencoders enable the extraction of meaningful features from high-dimensional data \cite{cae:2016}, while diffusion models support data enhancement tasks such as denoising and super-resolution \cite{cdm:2023, dpm:2015}. Deep Gaussian Processes provide a principled approach to uncertainty quantification in surrogate modeling \cite{deepgp:2020}. Additionally, transfer learning techniques, such as zero-shot hyperparameter transfer, generalize surrogate performance across tasks and model scales, offering efficient surrogates for deep learning computations \cite{mu:2021}.

\subsection{AI for Fault Detection}

Fault Detection or Anomaly Detection consists of automatically detecting abnormal behavior in monitoring data. Recent advances hold the potential to absorb different data and different levels of desired supervision according to the availability of manual labels. The tables~\ref{tab:ad} and \ref{tab:ad2} summarize this body of research.

\begin{table}[H]
\scriptsize
\setlength{\tabcolsep}{0.05cm}
\caption{Studies on AI for automatic fault detection in HPC between 2019 and 2022}
\label{tab:ad}
\begin{tabularx}{\textwidth}{>{\centering\arraybackslash}p{0.04\textwidth}>{\centering\arraybackslash}p{0.16\textwidth}>{\centering\arraybackslash}p{0.2\textwidth}>{\raggedright\arraybackslash}p{0.2\textwidth}>{\raggedright\arraybackslash}p{0.4\textwidth}}
\textbf{Ref} & \textbf{HPC Goal} & \textbf{AI Method} & \textbf{Data source} & \textbf{Outcomes}  \\
\cite{supversiedAD:2019}  & HPC Fault and Diagnostics & Supervised multi-class ML, Ensemble & Computer Failure Data Repository (CFDR) \cite{cfdr:2006} at NERSC from 2001 to 2006   &   90\% accuracy   \\ 
\cite{locateslowdown:2022}& Communication Fault  & Akaike Information Criterion, Analytic Hierarchy Process, Z-score, Median Absolute Deviation & 5 traces generated from three
HPC
systems: SMG2000 \cite{smg:2000}, AMG2013 \cite{asc:2013}, NAS BT \cite{nas:1994}. Traces generated with Score-P \cite{vihps:2022} & accurately identify execution
phases based on the detected communication patterns, it is capable
of identifying slow patterns within each phase, and determines
whether the latency is caused by late senders or receivers. These
inefficient patterns are categorized based on their severity and
complexity levels to help an analyst make decisions on where to
start the inspection. Open source. \\ 
\cite{anomalydetect:2022}  & HPC Fault    & Semi-supervised, Auto-encoder deep neural network & The data set combines sensor measurements coming
from a variety of HW sensors and program counters, information from the job scheduler (SLURM), and reports on the
system availability and status updates collected by Nagios \cite{nagios:2008}. Aggregated sampling of 5-minute interval: mean and standard deviation. 12874 samples. & anticipating the actual labels on average 45 minutes in advance. F-score is 85\% \\ 
\cite{ruad:2022} & HPC Fault & LSTM, Auto-encoder, reconstruction error & The
dataset contains 462 features. 10 months of operations. Sampling every 15 minutes and each
feature: minimum, maximum, average, and variance by Nagios \cite{nagios:2008}. & The proposed model architecture achieves the highest AUC
of 0.77 compared to 0.75, which is the highest AUC achieved by
the dense autoencoders. \\ 
\end{tabularx}

\end{table}

\begin{table}[H]
\scriptsize
\setlength{\tabcolsep}{0.05cm}
\caption{Studies on AI for automatic fault detection in HPC between 2023 and 2025}
\label{tab:ad2}
\begin{tabularx}{\textwidth}{>{\centering\arraybackslash}p{0.04\textwidth}>{\centering\arraybackslash}p{0.16\textwidth}>{\centering\arraybackslash}p{0.2\textwidth}>{\raggedright\arraybackslash}p{0.2\textwidth}>{\raggedright\arraybackslash}p{0.4\textwidth}}
\textbf{Ref} & \textbf{HPC Goal} & \textbf{AI Method} & \textbf{Data source} & \textbf{Outcomes}  \\
\cite{hdbscan:2023} & HPC Fault \& monitoring  & HDBSCAN, t-SNE \cite{tsne:2008} & 3 million entries or events and 89 features (Example: external temperature, PDUs measurement. Power from compute, storage, network, cooling units. Workload intensity, power usage effectiveness, ...) & Three phases methodology: raw data cleaning, ML-based processing and visualization \\
\cite{anomalydetect:2023}  & HPC Fault & GNNs, Binary classification & 31 months of data & Compared to SoTa Recurrent Unsupervised Anomaly Detection (RUAD) \cite{ruad:2022}. 1.5 hours ahead: AUC is 0.8826 against 0.8699 with RUAD. 72 hours ahead: 0.6219 against 0.5639 with RUAD \\ 
\cite{graafe:2024}      & HPC Fault    & Graph Convolutional Network, rack-level and room-level GNNs  & 31 months. 15-minute sampling frequency. It uses ExaMon \cite{examon:2023}, which aggregates a wide set of telemetry data. & It comprises three
main parts: The monitoring subsystem, the MLOps subsystem with Kubernetes, and the
anomaly the GRAAFE GNN model. It achieves an area under the curve (AUC) from 0.91 to 0.78, surpassing the state-of-the-art
(SoA), achieving AUC between 0.64 and 0.5. GRAAFE sustains the anomaly prediction for all the Marconi100
nodes every 120s, requiring an additional 30\% CPU resources and less than 5\% more RAM w.r.t. monitoring only. Open source. \\
\cite{diskfail:2024} & Disk Fault  & Ensemble, SMOTE Oversampling & Data from Self-Monitoring Analysis and Reporting Technology (SMART) of the disk. & compared with traditional methods, the F1-score of disk fault prediction is
improved by 6\%, and the model training time is also greatly reduced \\ 
{  \cite{nodesentry:2025}} & Node-level anomaly detection & HAC \cite{hac:2022}, Transformer, Mixture-of-Experts \cite{moe:2023}, Silhouettes\cite{silhouettes:1987} & Two real-world custom datasets & F1-score > 0.876; 0.560 better than baselines; 45.69\% training overhead reduction. Available open source. \\ 
{  \cite{refine:2025}} & Anomaly detection in contaminated data & VAE with iterative anomaly removal & Production HPC system logs, Hpas \cite{hpac:2019} & F1-score 0.88

\end{tabularx}

\end{table}

Most of the collected papers aim to detect a fault in multiple variables produced by the HPC system (network, CPU, power consumption, ...). Some others are focused on a particular element, such as communication traces \cite{locateslowdown:2022} or disk failures \cite{diskfail:2024}.

We identify the author model with different levels of supervision:
\begin{itemize}
    \item \textbf{Supervised multi-class classifier} \cite{supversiedAD:2019} allows to fit a model on previous data to detect coarse-grained anomalous patterns (hardware, software, network, human error). It relies on monitoring HPC logs and associated manual labels to tag the different types of errors.
    \item \textbf{Supervised binary-classifier} \cite{anomalydetect:2023, graafe:2024, diskfail:2024}.  This method trains a model to detect anomalous patterns with binary labels (``abnormal'' and ``normal'').
    \item \textbf{Semi-supervised} \cite{anomalydetect:2022} is used when a large amount of unlabeled data is available, with a small subset containing binary labels.
    \item \textbf{Offline Unsupervised Anomaly Detection} \cite{ruad:2022} is employed when large amounts of data without anomalies are available to learn before anomaly detection is deployed. No anomaly patterns are provided for learning, instead, it identifies statistical deviations from normal behavior.
    \item \textbf{Unsupervised} clustering algorithms enable easy visualization of time series by identifying patterns and normal data trends. DBSCAN-based clustering methods, which include a ``reject'' cluster for anomalies \cite{hdbscan:2023}, as well as online unsupervised anomaly detection techniques \cite{locateslowdown:2022}, are commonly used for this purpose.
\end{itemize}

Finally, some studies adopt hybrid approaches that combine different modes of supervision. For instance, \cite{ruad:2022} suggests switching between different levels of supervision to balance between high-human intervention with high-prediction quality and no labels, but lower precision. Others, such as \cite{anomalydetect:2023}, use an ensemble of two different AI models predicting at the same time and combine their advantages: one is self-supervised training on a large quantity of unlabeled data, and the second is a supervised model trained on a smaller quantity of data but labeled by a human expert.

Various data structures can be ingested by AI models. Recurrent models, such as Long Short-Term Memory  \cite{ruad:2022} networks, have proven to be well-suited for extracting information from temporal data. More recently, Graph Neural Networks \cite{graafe:2024} have demonstrated success in processing graph data by learning from different interconnected components within a network. GNNs have shown superior performance in detecting anomalies in computer networks compared to traditional neural networks that handle tabular data, as they effectively capture neighborhood contextual information between components.

\subsection{LM for HPC scripting/coding automation}
{ Language Model (LM)} may be used to automate various aspects of information systems such as virtual assistance, search engines, and document summarization. Those approaches are especially useful in the highly technical and scientific context of the HPC. Table~\ref{tab:nlpauto} summarizes recent research in this field.

\begin{table}
\scriptsize
\setlength{\tabcolsep}{0.05cm}
\caption{Studies on { Language Model (LM)} for HPC operations automation}
\label{tab:nlpauto}
\begin{tabularx}{\textwidth}{>{\centering\arraybackslash}p{0.04\textwidth}>{\centering\arraybackslash}p{0.16\textwidth}>{\centering\arraybackslash}p{0.2\textwidth}>{\raggedright\arraybackslash}p{0.2\textwidth}>{\raggedright\arraybackslash}p{0.4\textwidth}}
\textbf{Ref} & \textbf{HPC Goal} & \textbf{AI Method} & \textbf{Data source} & \textbf{Outcomes}  \\
\cite{litterworkflow:2020}                           & Summarization of scientific papers on exascale computing topic                                                                                                              & Big Data workflow, T-SNE \cite{tsne:2008}, Clustering    & Scopus \cite{scopus:2018} & Methodology with three-stage approach: (1) the challenges and opportunities of exascale based on various landmark studies are identified. (2) Data-driven techniques are used to analyze the large collection of related academic literature. (3) Eight research themes are identified                                                                                                                        \\
\cite{bearicade:2021}                                & Supervised NN for cyber-attack                                                                                                                                                        & Neural Network, Anomaly Detection        & 30,000 activities from the users. They contain the occurrence of the command ran and the occurrences &  toolbox of reusable Ansible roles and playbooks to configure a cluster software environment described by the freely available OpenHPC recipes. Open source.                                                                                                                               \\
\cite{nlu:2022}                                      &{ LM}to design HPC workflow                                                                                                                                                  & { LM}for code generation HISyn \cite{hisyn:2020}     & HPC-Fair \cite{hpcfair:2021} methodology. 60 custom Natural Language queries: 17 data manipulation, 25 ontology-interactive queries, 18 combined queries. &    80\%
synthesis accuracy. The interactive design
achieves 95\% accuracy for extracting ontology information
inside the NL query                                                                                                                 \\
\cite{hpcgpt:2023} & Managing AI models and datasets for HPC, and data
race detection & LLaMA-based model  & DataRaceBench V1.4.0 \cite{datarace:2023} containing 177
C/C++ test programs and 166 Fortran test programs. Among these,
88 C/C++ and 84 Fortran test cases exhibit data races, while 89
C/C++ and 82 Fortran test cases are free from data races.  & The proposed model exhibits the ability to retrieve and extract HPC-related datasets and models based on human expressions. Good performance
compared to existing data-race detectors. \\
\cite{codex:2023}                                    & Automatic parallel code writing with diverse programming models, and diverse languages                                                         & OpenAI Codex \cite{codex}, GitHub Copilot  \cite{copilot}                 & Six fundamental kernels in HPC: HPC AXPY, GEMV, GEMM, SpMV, Jacobi Stencil, CG &    an average of novice level, for the languages and kernels tested.  {  There is} a slightly higher quality in the automatically generated C++ and Python languages rather than in Julia and Fortran. \\ 
\cite{firewall:2024}                                 & ML Classifier for Firewall Rule                                                                                                                                                          & KNN, Neural Network                     & 4.6 billion samples with 289,320 instances of identified firewall rule-policy errors. 10 features: source IP (4 parts), destination IP (4 parts), source
port, and destination port &  The results indicate that the KNN and NN models exhibited
an accuracy of 97\%. Additional training and feature refinement led to even better improvements, increasing the accuracy to 98\%    \\ 
\cite{llm:2024} & OpenACC, OpenMP automatic test-suite generation & Retrieval-Augmented Generation \cite{rag:2020}, ChatGPT \cite{gpt:2020}, DeepSeek-Coder \cite{deepseek:2024},  fine-tuning, prompt engineering     & 1335
training examples, 351 test examples & Deepseek-Coder-33b-Instruct produced the most passing
tests of any LLM: 170/351   \\ 
{  \cite{synmarco:2025}} & Parallel code generation & Multi-agent LLM: code generation, performance estimation, web-search  & 75 LeetCode problems & 14.6\% average runtime reduction compared to Claude 3.5 Sonnet
\end{tabularx}
\end{table}

{  Below, the literature is summarized according to the different goals addressed in the research papers.}

\begin{itemize}
\item \textbf{Information retrieval in big data}. The HPC application and infrastructures are gaining in complexity due to the heterogeneity of platforms. AI has been useful in detecting the trends in the thousands of research papers published every year \cite{litterworkflow:2020}.
\item \textbf{Cybersecurity Script Generation}. HPC environments, accessible through public networks, are often prime targets for a wide range of cyberattacks. To improve security, machine learning algorithms \cite{bearicade:2021, firewall:2024} have been proposed to design and assist HPC administrators in writing firewall rules and scripts. They can analyze historical logs to identify patterns and vulnerabilities, enabling them to build new rules. 
\item \textbf{Parallel code generation for HPC operations and coding}. This approach involves automatically designing software or assisting HPC developers through automatic code generation \cite{nlu:2022, llm:2024}. The advancement of these methodologies is particularly beneficial in the context of complex systems, driven by the recent development of diverse accelerators and their intricacies. 
\end{itemize}

\section{AI for HPC Summary}
\label{sec:sum}

 A deeper literature analysis uncovers interdependencies between these objectives. AI techniques for performance estimation can be applied in various contexts such as workload scheduling, anomaly detection, and performance optimization. These applications benefit both HPC users (through optimization, performance estimation, and AI surrogates) and HPC administrators (through anomaly detection, automation of operations, and workload scheduling). By leveraging AI, both users and administrators can achieve more efficient, reliable, and effective HPC systems.

The future of High-Performance Computing (HPC) is being shaped by a paradigm shift toward heterogeneity, encompassing a broader variety of accelerators and CPU architectures. This trend makes efficient scheduling and diverse computation acceleration more critical than ever before.

{

\subsection{Category-Specific Challenges.}

While several overarching barriers affect AI integration in HPC, each application domain presents unique challenges:
 
\begin{itemize}
    \item \textbf{Performance estimation.} Model generalizability remains a major difficulty due to hardware and workload heterogeneity—models trained on one system often underperform on others. This limits their portability across HPC centers. Finally, a lack of standards for HPC workloads makes estimation and comparison difficult. This is not true in the pure deep learning task , due to their nature to be a directed acyclic graph of tensor computation, some estimator may be designed \cite{colossal:2023}. A universal prediction of software execution time is theoretically impossible in the general case, as it depends on undecidable properties of programs (e.g., the halting problem), so practical performance estimation must rely on statistical models and empirical domain-specific measurement. 
    \item \textbf{Performance Optimization.} Several works motivate new optimization designs by pointing out the scalability limits of prior optimizers. Another open challenge is the reuse of knowledge across tasks, addressed in \cite{multitasktuning:2023} through transfer learning and multitasking, where insights from one optimization process accelerate learning on related tasks. Finally, a third challenge lies in the growing complexity of the parameter space: optimizers must handle high-dimensional search spaces combining system- and application-level settings \cite{bayes:2020}.
    \item \textbf{Scheduling.} First, scalability: many approaches, such as RL-based schedulers \cite{DRAS:2022, garlsched:2022} show promising results in simulations, but have not been demonstrated at scale in production systems. Second, trust remains limited: ML-based schedulers often act as black boxes, making them harder to explain and less reliable than heuristics like SLURM. Finally, schedulers must balance multiple and sometimes conflicting objectives—such as makespan, queue waiting time, fairness, and energy cost \cite{tardis:2025}.
    \item \textbf{Fault detection.} Techniques often rely on black-box models, raising concerns about interpretability and trustworthiness \cite{synfaulsched:2013}. Providing explainable anomaly scores and actionable diagnostics remains an open problem. 
    \item \textbf{Surrogate modeling.} While effective in reducing computational cost, surrogates often lack systematic approaches to quantify uncertainty in high-stakes simulations \cite{bayes:2020, multitasktuning:2023}. Moreover, there is no consensus on standardized datasets or evaluation metrics, limiting transfer learning techniques and cross-study comparability.
    \item \textbf{Code and scripting automation.} General-purpose LLMs exist and enable comparison with research works in the HPC context. However, papers in this area are rare and pursue heterogeneous objectives (e.g., parallel code generation \cite{synmarco:2025}, testing \cite{codex:2023}, bug detection \cite{hpcgpt:2023}, ...) without shared benchmarks across studies. There is no documented evidence of production-level deployment to date, and the reliance on simplified tasks (e.g., LeetCode problem sets \cite{synmarco:2025}) does not reflect the skills required for developing and maintaining large-scale HPC codebases in production
\end{itemize}

These domain-specific issues demonstrate that generalized integration of AI into HPC remains challenging unless each category first resolves its internal limitations.
}

{

\subsection{Platform-related Challenges}

\textbf{Challenge: Accelerator heterogeneity.}

While the majority of GFLOPS in modern HPC systems are delivered by AI accelerators, primarily GPUs with SIMT architectures, the challenge of scheduling across heterogeneous AI accelerator types remains largely unaddressed. Recent studies on alternative architectures, such as Graphcore IPUs with MIMD architecture, have demonstrated substantial performance gains for specific workloads, with speedups reaching several factors over traditional GPUs \cite{arcelin:2023, ipu:2023}. Despite this, current HPC systems often place the burden of accelerator selection, code adaptation, and performance experimentation entirely on users. In contrast, AI-based schedulers, similar to those emerging in cloud environments \cite{oiko:2023}, could provide intelligent guidance by selecting the most suitable accelerator dynamically, leveraging workload characteristics, real-time system states, and historical performance profiles.
}

\textbf{Challenge: MLops of AI-based scheduler.}

``AI for scheduling'' remains the most active subfield in addressing the broader research question, ``What can AI do for HPC systems?''. Furthermore, most AI models for performance estimation are designed specifically to assist scheduling. The complexity of AI workloads—due to architectural diversity, computational constraints, and workload variability—requires advanced scheduling techniques. Efficient resource allocation in AI infrastructures increasingly depends on intelligent workload characterization and predictive models that can adapt to evolving usage patterns.

Literature reviewed in Table~\ref{tab:sched}, \ref{tab:sched2}, \ref{tab:sched3} shows that heuristic-based scheduling often leads to sub-optimal decisions compared to AI-based schedulers, with significant potential for cost and energy savings. However, AI-based schedulers have not been implemented in MLOps pipelines, primarily due to challenges such as the lack of standardized benchmarks, the complexity of integrating learning models into real-time scheduling systems, concerns about model interpretability and trust, and the overhead of maintaining and retraining AI models in dynamic production environments.

\textbf{Challenge: Increasingly distributed workflows.}

The growing demand for heterogeneous computing is enabling new forms of inter-center collaboration, with the goal of maximizing accelerator utilization by intelligently matching user workloads to the most suitable hardware resources. Recent research emphasizes the development of heterogeneous computing networks that go beyond isolated HPC centers. Initiatives such as Grid'5000 \cite{grid5000} and SLICES-RI \cite{slices:2022} are pioneering the creation of a next-generation computing ecosystem that supports distributed HPC, AI acceleration, advanced networking, and shared storage infrastructure. SLICES-RI, a collaboration among 16 European countries, provides access to a diverse array of resources—including HPC nodes, AI accelerators, IoT devices, and networking equipment—designed to facilitate geographically distributed workflows and multidisciplinary research.

This shift brings both opportunities and challenges for scheduling, programming, and detecting anomalies in multi-site HPC systems, HPC/IoT integration, and bridging HPC with cloud platforms. As a result, future AI for HPC research must consider this increasing heterogeneity, not only in terms of processor capabilities, but also in communication bandwidth and latency between distributed processing units.

\subsection{ AI trends  }

AI methods in the context of HPC are rapidly evolving, with deep learning approaches making significant strides. Deep Gaussian Processes (DeepG) have proven effective for quantifying uncertainties related to performance optimization \cite{multitasktuning:2023} and surrogate modeling \cite{deepgp:2020}, enabling robust decision-making in high-dimensional nonlinear systems. Graph Neural Networks (GNNs) have emerged as a powerful tool for capturing computer information and their interdependencies in networked machines \cite{anomalydetect:2023, graafe:2024}. Additionally, innovative applications are being developed that leverage natural language processing to streamline the creation and scaling of complex parallel code \cite{nlu:2022, codex:2023, llm:2024}, enhance cybersecurity scripts \cite{firewall:2024}. Finally, actor-critic reinforcement learning has consistently outperformed heuristic approaches, as demonstrated by the studies in tables \ref{tab:sched}, \ref{tab:sched2}, \ref{tab:sched3} despite the challenge of handling large-scale action spaces required for assigning jobs to resources. Achieving strong results with a large action space has been made possible through techniques such as action masking \cite{smdp:2023} or by employing reinforcement learning to select optimal heuristics from a predefined pool rather than making direct predictions \cite{DAG:2024}.

AI methods in HPC are advancing rapidly, with several key trends emerging. Natural Language Processing is being explored for a variety of HPC applications, such as generating kernel codes, creating tests, and orchestrating complex workflows \cite{llm:2024, codex:2023, nlu:2022}. Recent research focuses on replacing large LLMs with smaller, more efficient models for inference. While the autoregressive nature of LLMs presents challenges for parallelizing inference, studies have demonstrated a 2x speedup when substituting large models with smaller ones in specific scenarios \cite{specinfer:2024}. This highlights the potential for optimizing inference procedures to achieve greater efficiency.

Advances in AI have significantly improved the ability to interpret unstructured data, enabling better contextual understanding. Graph Neural Networks (GNNs) are emerging as a cutting-edge approach in anomaly detection, effectively capturing dependencies in computer networks and outperforming traditional models \cite{graafe:2024, anomalydetect:2023}. LLMs are also proving invaluable for consuming and generating code. Tools like DeepSeek-Coder \cite{llm:2024} push the boundaries toward Artificial General Intelligence (AGI), automating the creation of code and tests via intuitive prompting.

Alongside model development, research is increasingly emphasizing Machine Learning Operations (MLOps) for HPC. These workflows focus on deploying AI predictions effectively and include tasks such as retraining models in production environments, conducting hyperparameter searches, and monitoring input data and predictions \cite{hdbscan:2023, graafe:2024}. Additionally, methodological efforts aim to establish standardization in AI workflows \cite{hpcfair:2021} and facilitate the collection and organization of scientific literature \cite{litterworkflow:2020}.

Reinforcement Learning has consistently demonstrated superior performance in task scheduling when compared to traditional heuristics, such as First-Come-First-Serve with backfilling queues \cite{garlsched:2022, DRAS:2022, rlschedc:2021}. Despite this, heuristics remain a widely used approach in many studies due to their ability to explore large decision spaces and produce relevant decisions without requiring prior interactions or extensive training \cite{eelas:2023, elasticflow:2023, pochelu:2022}. These two families of methods—RL and heuristics—continue to coexist, with hybrid approaches seeking to combine the strengths of both to achieve more effective and adaptable solutions \cite{DAG:2024}.

The future of AI in HPC likely involves the convergence of techniques like scripting/coding generation, scalable reinforcement learning, uncertainty estimation, and graph-based information extraction. Integrating these methods could significantly enhance job characterization and control, leading to more efficient and sustainable computing infrastructures.

{ 

\subsection{ Meta-Challenges in the Literature }

}

HPC centers worldwide often operate with distinct priorities and goals, making it difficult to design universal AI solutions. For example, job scheduling may be formulated to minimize average job completion time, ensure fairness among users, reduce energy consumption, or lower infrastructure costs \cite{garlsched:2022}. Similarly, anomaly detection may target different priorities, such as optimizing F-score metrics \cite{anomalydetect:2022}, enhancing explainability \cite{ruad:2022}, and diagnosing root causes \cite{locateslowdown:2022}. These diverse priorities necessitate that administrators can customize AI models to balance priorities and priority shifts.

AI solutions for HPC often rely on datasets derived from hardware sensors, job schedulers, and system logs. However, these datasets are frequently unique to specific HPC centers, complicating efforts to develop AI models that generalize across different environments \cite{supversiedAD:2019}. Increasing the size and diversity of data inputs can improve model performance, but also risks overfitting to the idiosyncrasies of a single HPC center \cite{anomalydetect:2022}. The lack of standardized data sources further hinders the creation of versatile AI solutions applicable across heterogeneous systems.

Privacy concerns related to HPC user activity present additional barriers to sharing datasets. While publicly available datasets \cite{cfdr:2006, swfarchive:2014, lublin:2003} exist, they often fail to capture recent trends, such as GPU-accelerated workloads and AI applications. Many AI models are trained using proprietary data or simulators, limiting the reproducibility and collaborative advancement of AI research in HPC \cite{ruad:2022}.

The challenge of generalization also affects AI models trained on homogeneous HPC systems, as they often fail to perform well in heterogeneous environments with varied nodes and accelerators \cite{DRAS:2022}. Similarly, HPC traces from one center may not accurately represent the demand patterns or scheduling requirements of other centers. For instance, centers with high AI workloads face high accelerator demand compared to those focused on traditional CPU-based tasks.

While reinforcement learning has demonstrated its potential in HPC scheduling \cite{garlsched:2022}, fully integrated workflows remain rare. These workflows must encompass diverse components, such as retraining mechanisms to address data shifts, hyperparameter optimization, administrator-facing dashboards, and explainability of RL-based scheduling decisions. The development of such comprehensive solutions is an ongoing challenge in advancing AI applications for HPC \cite{DAG:2024}.

{   In addition to technical challenges, the literature itself presents difficulties in synthesis and comparison. The level of experimental detail varies significantly across papers. Some studies offer comprehensive performance metrics, while others omit critical information such as dataset sources, validation protocols, or evaluation baselines. For instance, several methods report no dataset at all, either because no historical data was needed, the method remains untested, or the authors did not disclose it. Furthermore, the use of heterogeneous evaluation metrics (e.g., speedup factors, percentage reductions, accuracy scores) without standardization complicates direct comparison and meta-analysis. These limitations hinder the establishment of benchmark-driven conclusions and highlight the need for better reporting practices in the HPC field. 
}

To address these challenges, standardizing data collection and developing common frameworks for HPC datasets are crucial. Initiatives aimed at creating shared resources can facilitate collaboration and improve the generalizability of AI solutions. Additionally, customizable AI methods that align with the specific goals of individual HPC systems hold promise for overcoming the limitations posed by diverse objectives and environments \cite{hpcfair:2021}.

\section{Integration of AI models into HPC systems}
\label{sec:integr}

This section outlines the LLM-OS concept, its potential to integrate multiple ``AI for HPC'' models seen in previous sections (performance optimizers, surrogates, schedulers, fault detectors, and code generators) into a unified framework, and the key challenges it presents.

\subsection{LLM-based Operating Systems}
\label{sec:llmos}

Traditional operating systems (OS) manage hardware and software resources to ensure efficient computing. With the emergence of large language models (LLMs), a new paradigm is being proposed in which LLMs operate not just as applications but as part of the core OS—managing memory, processes, and external tools \cite{aios:2024, memgpt:2024, llmos:2023}. This concept transforms OS architecture by embedding reasoning, task planning, and contextual awareness directly into the system’s core.

Modern OSs typically adopt a layered and modular architecture, comprising components such as networking, storage, peripheral device support, interpreters, and development toolkits. These structures promote modularity, extensibility, and maintainability—properties that are equally desirable when integrating LLMs into the OS kernel.

Recent advances in AI technologies, including language reasoning, computer vision, and speech recognition, have significantly expanded human-computer interaction. Voice assistants like Apple’s Siri\footnote{\url{https://www.apple.com/siri/}} and Microsoft’s Voice Access\footnote{\url{https://www.microsoft.com/en-us/windows/tips/voice-access}} demonstrate early forms of natural language control. However, these systems typically operate at the application level. In contrast, LLM-based OS architectures propose a deeper integration of AI at the core system level, enabling AI-native process management, tool orchestration, and interactive user experiences.

LLM-based Operating System (LLM-OS) architectures leverage recent advancements in large language models (LLMs) to mimic several OS modules. They can integrate other agents (the applications running above the OS). These systems aim to unify and orchestrate previously trained AI agents while maintaining a natural language interface. Table~\ref{tab:aios_comparison} highlights the mapping between traditional OS components and those of a typical LLM-OS, highlighting the feasibility and modularity of LLM-OS with the advantage of being powered by AI for further flexibility and adaptability.
\begin{table}[H]
\centering
\caption{Mapping between traditional OS elements and LLM-OS elements reported from \cite{llmos:2023}.}
\begin{tabular}{|l|l|}
\hline
\textbf{Traditional OS} & \textbf{LLM-OS} \\
\hline
Kernel & Core LLM-OS \\
Memory & Context Window \\
Memory Management & Context Selection and Management \\
File System & Retrieval Storage \cite{memgpt:2024} \\
Applications/Libraries & Software Agents \\
Driver/API & Prompts/Instructions \\
OS SDK & LLM OS SDK \cite{llmos:2023, aios:2024} \\
User Interface & Natural Language Prompt \\
Applications & Agent Applications \\
\hline
\end{tabular}
\label{tab:aios_comparison}
\end{table}

MemGPT \cite{memgpt:2024} lays foundational work for LLM-OS memory management. It adopts a hierarchical memory model to preserve the most important information (user intent, user profile, current in-context document), inspired by traditional OS principles allowed through Retrieved-Augmented Generator \cite{rag:2020}, enabling long-term context management through:
\begin{itemize}
    \item \textbf{Main context:} a limited but fast in-context window.
    \item \textbf{External context:} long-term memory, typically stored on disk or in external databases \cite{rag:2020}.
\end{itemize}
This abstraction allows LLMs to handle extended document analysis and maintain coherent multi-session interactions by dynamically swapping memory, similar to virtual memory in classic OS design.

Several models have been proposed \cite{llmos:2023,aios:2024,aiossurvey:2024} that extend this concept further by proposing a system where LLMs function as the OS kernel. Within this framework, both software and AI agents act as applications, dynamically orchestrated by the LLM core based on user input, available resources, and context.

A key feature of LLM-OS architectures is their ability to interpret a wide range of diverse user goals, decomposing them into a chain of thought \cite{cot:2022} that translates high-level intents into a sequence of simpler tasks, such as API calls executed by agents (or AI or software agents) \cite{memgpt:2024, openagi:2023}. An illustrative example from \cite{aios:2024} involves a user planning a trip, which is broken down into software-level actions (e.g., uploading a photo ID, processing payments, calendar planning) and calls to specialized agents (e.g., retrieving user preferences, recommending flights and hotels).

These systems are described as containing the following components:
\begin{itemize} 
\item LLM-OS: it interprets user intent and decomposes high-level goals into structured API calls. It requires access to agent APIs and, when needed, selects API calls demonstrations \cite{tptu:2024} that align user instructions with corresponding API calls.
\item AI Agents: domain-specific models designed for tasks such as reasoning, data analysis, decision-making, and content generation. 
\item Software Agents: it is a wrapper for software, including desktop tools, software development toolkits (SDKs), interpreters, or simulators. 
\item Agent Caller: a scheduling and coordination module (also referred to as the AI-OS Kernel in \cite{aios:2024}). It manages agent execution and enforces access control. This component ensures safe operation by preventing unauthorized access and managing computational resources. It utilizes metadata such as resource requirements (e.g., compute time, memory usage, GPU availability), agent dependencies with versioning, and access rights to efficiently and safely schedule agent execution on the underlying hardware.
\end{itemize}

\begin{figure}[h!]
\centering
\includegraphics[width=\columnwidth]{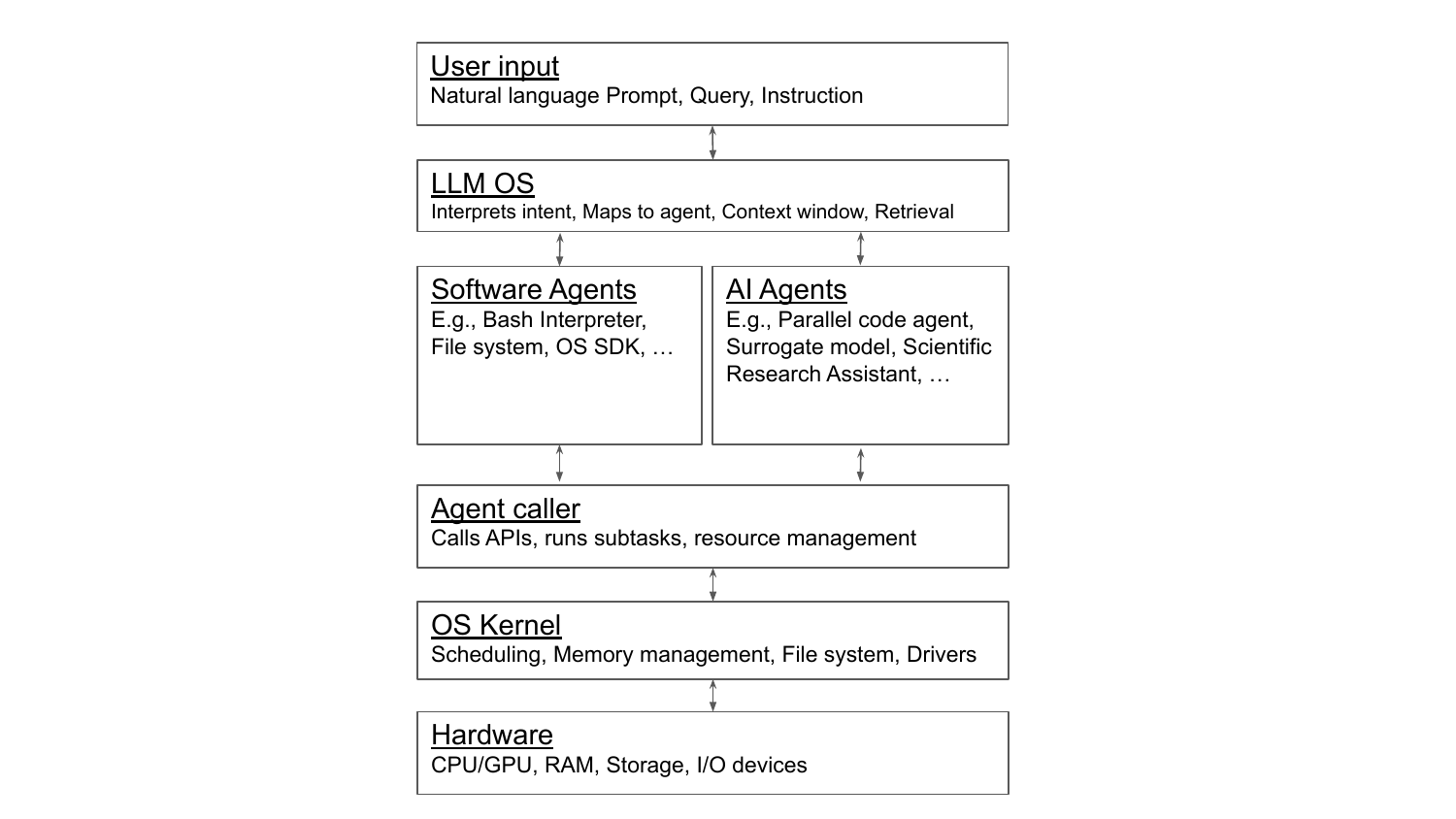}
\caption{Dataflow stack of LLM-OS inference from user intent to hardware}
\label{fig:llm_os_stack}
\end{figure}

The figure~\ref{fig:llm_os_stack} illustrates the dataflow of the inference of a standard LLM-OS described by \cite{aios:2024, aiossurvey:2024}.

\subsection{Opportunities}
\label{sec:synergy}

The true potential of AI in HPC lies not only in isolated innovations but in the strategic integration of complementary specialized AI models. LLM-OS interfaces provide a natural language layer that can unify control over these AI subsystems and collect and select the most relevant information generated by the agents. 

\begin{figure}[h!]
\centering
\includegraphics[width=\columnwidth]{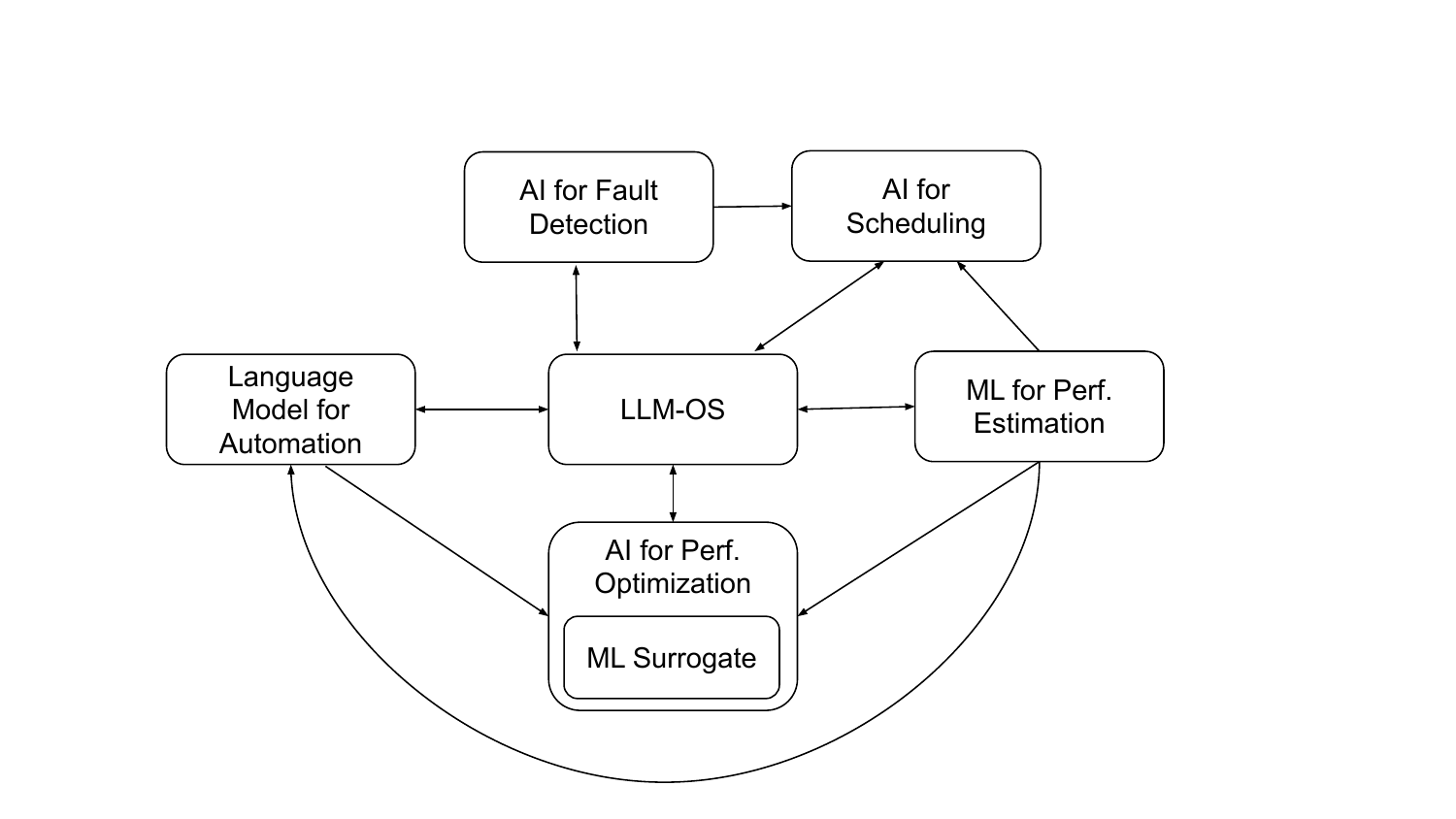}
\caption{Synergies between the seven potential ``AI for HPC'' sub-domains. Each node represents a specific AI application area within HPC.The edges represent direct potential interactions and the information flows. Concrete examples of these synergies are discussed in the subsequent text.}

\label{fig:synergies}
\end{figure}

Among the seven AI for HPC models and LLM-OS (not yet deployed for HPC), multiple synergies are identified and summarized in Figure~\ref{fig:synergies}.

{

The conceptual links between subtopics are presented below with illustrative works. Software maturity is also discussed.

\begin{itemize}
\item \textbf{AI Performance Estimation → Scheduling}. Predicting job runtime, queue wait time, or power consumption is a proven way to improve HPC scheduling. By contrast, default systems such as SLURM often assume nothing about incoming jobs and rely on simple First-Come-First-Serve rules. Many workflows now combine estimators with schedulers. Some use supervised prediction to guide standard schedulers \cite{ear:2020, sensetimechara:2021}. Others link supervised estimators with RL schedulers \cite{rlschert:2021}. Hybrid designs combine supervised estimators with custom optimization policies \cite{tardis:2025}. In highly heterogeneous clusters, schedulers are often paired with recommender systems to guide job placement \cite{geophy:2024, oiko:2023}. A different line of work explores pure RL schedulers \cite{rlschedc:2021, DRAS:2022, DAG:2024}, which do not rely on explicit estimation. They can learn powerful strategies, but are usually less interpretable and remain mostly experimental. In practice, only the EAR system \cite{ear:2020} demonstrates an estimator combined with a scheduler in production. Its plugin, \textit{earplug.so}, integrates with SLURM and reduces power consumption using a supervised linear regression model \cite{linearear:2011}.
\item \textbf{AI Performance Estimation → Performance Optimization}.Some performance optimization frameworks reduce sampling costs by incorporating data-driven performance estimation. For instance, Bayesian optimization \cite{bayes:2020, multitasktuning:2023, autotuning:2024} typically involves three phases: sampling, modeling, and searching. Transfer learning \cite{multitasktuning:2023} can further accelerate this process by reusing knowledge from related tasks, thereby lowering the number of expensive evaluations needed. Among optimization systems deployed in production (Merlin \cite{merlin:2022}, Alpa \cite{alpa:2022}, and Colossal-AI \cite{colossal:2023}), only Colossal-AI integrates an AI-based estimation component, Colossal-Auto \cite{syncolossalauto:2023}, which performs prediction prior to optimization.
\item \textbf{LM Code/Scripting Automation → Performance Optimization} The semantic link is straightforward between code/script automation in the HPC context and optimization. Most of the collected papers in this category aim at automation to best use resource and achieve performance, accelerate converge, or scalability. However, none of those papers’ experiments have been assessed on a large scale.
\item \textbf{Language Model → Performance Estimation + Performance Optimization}. LLM-based systems, such as MARCO \cite{synmarco:2025}, perform joint tasks by generating optimized code/scripts while also providing performance feedback. These models can estimate runtime from user-submitted code and constraints, and iteratively refine the code for better efficiency, thus bridging performance estimation and optimization in a unified automation loop.
\item \textbf{AI for Fault Detection → Scheduling}.To improve the resilience of HPC systems, the fault detection mechanisms can be integrated into scheduling. A proactive approach can prevent workload failures by combining three key software components: monitoring, early fault detection, 
and preemption through task re-scheduling to healthy nodes. Such proactive fault-tolerant systems are well established in cloud computing \cite{synfaultsched:2024, synfaulsched:2013},  where virtualization facilitates VM migration (in case of hardware failure) or scaling (in case of resource exhaustion). Comparable frameworks in the HPC context remain rare in the literature.
\item \textbf{ML Surrogate $\subset$ ML Performance Optimization}. Surrogate modeling is a specific strategy within the broader field of performance optimization using machine learning. These models aim to approximate expensive computations in HPC workloads, enabling rapid evaluations during tuning or design space exploration. 
\item \textbf{LLM-OS and others} As outlined in Section~\ref{sec:llmos}, LLM-OS layers can serve as orchestration points of specialized AI models and components through shared knowledge. However, their application within HPC systems is still purely conceptual, as no empirical deployment or experimentation has yet been reported.
\end{itemize}

Despite these observed synergies, a major challenge remains: the lack of integration and standardization across AI components for HPC systems. As highlighted in the survey by \cite{survworkflow:2024}, the absence of standardized APIs, data formats, and interfaces impedes the seamless composition of AI models developed independently. This fragmentation undermines the reusability of components, the interface between user workflows and AI tools, the portability of AI solutions across domains, the reproducibility of experiments, and the ability to perform fair comparisons between models. While initiatives such as the FAIR principles, which promote workflows that are Findable, Accessible, Interoperable, and Reusable, have been proposed \cite{fairworkflow:2022}, their adoption remains limited. Consequently, even when technical synergies exist, practical integration is hard to achieve.

Furthermore, several conceptual links between AI tasks remain underexplored or missing. For instance, joint HPC code generation and runtime-time prediction LLM feedback loops are promising avenues but lack representation in current literature \cite{synmarco:2025}. Some research, like SYNDICATE \cite{syntogether:2023}, proposes tightly coupled optimization of scheduling and execution planning, but its scope is limited to deep learning communication collectives, making generalization to all HPC workloads impossible.


}

A potential ``ML for Performance Estimation'' agent capable of providing accurate time and memory usage predictions could serve multiple roles. It could inform users about whether it is worthwhile to launch a job, assist the scheduler in deciding when and where to run jobs based on resource availability and expected load, and guide the code design and tuning process. These estimations could also be surfaced through LLM interfaces, enabling users to evaluate the cost-benefit ratio of launching a computing task.

However, a performance estimation agent is not always necessary to inform other AI models. For example, RL-based performance optimization often represents a model-free approach that bypasses prior estimation, learning instead through direct interaction with the environment. Similarly, some scheduling strategies or code tuning heuristics operate effectively using static rules or hardware-aware heuristics (e.g., cache hierarchies), rather than explicit performance predictions. Even in LM-based automation for code optimization, performance estimators are rarely used as intermediate models—most approaches rely on real experimentations or hardcoded performance rules, though a hybrid system where estimators are used as reward signals for RL-based { Language Model (LM)} is a promising future direction.

The three subfields ``ML Surrogate Models'', ``AI for Performance Optimization'', and ``{ Language Model} for (Parallel Code) Automation'', share a common goal: accelerating computation, albeit through different mechanisms. Surrogates aim to replace portions of numerical simulations (often with human-in-the-loop placement), optimizers fine-tune input parameters to maximize performance, and { Language Model} agents automatically generate or restructure code. Their functionalities are naturally interconnected.

For example, { Language Model} could generate code that is then further optimized by a performance optimizer (e.g., tuning message sizes, process-core mappings, or shared memory allocation). Conversely, profiling results from performance optimization tools could inform a { Language Model} code generator to revise suboptimal code segments. Furthermore, { Language Model} agents could identify recurring patterns in code (e.g., redundant functions or loops) and trigger supervised ML integration or surrogate replacement automatically.

ML surrogate models, designed to approximate costly simulations, also integrate with performance optimization workflows, since both aim to reduce the runtime or resource consumption of HPC applications. These interconnections support an emerging architecture of AI for HPC: a system of feedback-driven models that share data and reinforce one another. Whether through shared training pipelines, real-time inference loops, or coordination via LLM-based interfaces, these models amplify their impact when deployed in concert.

At the level of a single HPC job, similar synergies appear. Surrogate models can be embedded within applications to replace expensive computations, while performance optimizers can adapt parameters dynamically, informed by surrogate feedback. This allows faster iteration and reduced computational overhead in tuning complex workloads.

At first glance, several orthogonalities are also observed between AI models that necessitate no direct communication. In particular, models acting \textit{within} jobs (such as performance optimizers and { Language Model} automation) address goals related to code execution and tuning, whereas models acting \textit{outside} jobs (such as AI-based schedulers and fault detectors) operate at the system orchestration level. These models solve different problems and may evolve independently.  

However, orthogonality in functionality does not imply isolation. Many of these models may consume overlapping contextual information, such as user behavior patterns, historical performance data, or hardware topology, and contribute outputs that are useful for other agents. A unifying, higher-level agent such as an LLM-OS can serve as an integration layer, allowing models to interoperate indirectly through shared knowledge representations. For example, a fault detection model might identify a degraded node and update a global status register. While the performance optimizer embedded could receive hints or modified execution recommendations from the LLM-OS, which aggregates infrastructure context. Similarly, a scheduler may use user or application profiles gathered during past optimization runs to make future scheduling decisions, even if those agents never directly interact.

This results in a complex flow of information, where some AI agents communicate directly, exchanging data and API calls, while others interact only indirectly by accessing and contributing to shared knowledge bases about users, applications, and hardware. The LLM-OS plays a central role in this architecture.

\subsection{Challenges}

\textbf{LLM-OS Reasoning.}

A core challenge in LLM-OS design lies in enabling effective interaction between language models and agents via API calls. This process demands a combination of Chain-of-Thought (CoT) reasoning \cite{cot:2022}, the ability to plan sequences of API calls, and the practical utilization of external tools. Decomposing user prompts into subtasks using CoT has demonstrated strong performance in mathematical and symbolic reasoning tasks \cite{reasoning:2023}; however, it does not consistently improve outcomes across broader domains relevant to HPC, such as content generation, context-aware question answering, or tasks requiring expert-level knowledge.

Moreover, studies show that LLMs still struggle with planning and reasoning tasks that are trivial for humans \cite{reasoning:2023}. This highlights the need for more extensive fine-tuning \cite{tptu:2024}, incorporating demonstrations of effective API usage and multi-step planning. For instance, evaluations on structured reasoning benchmarks reveal that even state-of-the-art models GPT-4 fail in abstract, multi-step planning tasks \cite{blocksword:2024}.

\textbf{LLM interactions.}

Each AI agent \cite{multiagent:2024} must align its assigned tasks with overarching system objectives while also integrating contextual information from other agents. Ensuring consistency in shared goals and effectively leveraging distributed contextual knowledge remains a significant challenge in multi-agent LLM-OS environments. Recent research \cite{llm2llm:2024} shows that when two LLMs engage in cooperative problem-solving via iterative question-answer exchanges, the original objective often degrades over time. Lengthy prompt exchanges can cause the focus to drift, leading to less relevant or incoherent responses. However, emerging findings suggest that state-of-the-art models, such as GPT-4, are more resilient in maintaining coherence and goal alignment during extended interactions.

Effective communication in traditional HPC systems, between operating systems, applications, and orchestration layers, relies on standardized protocols and domain-specific languages (DSL) such as Terraform \cite{terra:2022}. Similarly, Large Language Model Operating Systems (LLM-OS) and AI agents are beginning to adopt standards for data and instruction exchange \cite{llmos:2023}. However, striking a balance between the expressive flexibility of natural language and the structural rigidity of DSLs remains a significant challenge. The inherently unstructured nature of LLM interactions complicates traditional security and compatibility frameworks, requiring robust oversight, validation mechanisms, and carefully designed interfaces to ensure smooth integration across heterogeneous platforms and versions.

\textbf{System Integration.}

In contrast to monolithic AI models, integrating AI models for HPC requires coordination between all aspects: updating the callable software/AI APIs, software, application data, AI model versioning and repository, and triggering of update of AI models and software when infrastructure changes.

The interaction between AI agents is far from easy. Even if multiple agents aim for the same goal (e.g., HPC job scheduling), their operational requirements may diverge significantly. For example, the collected tables of this article show that some AI-enhanced schedulers may require real logs, some synthetics from HPC digital twin, some consume the prediction of a performance estimator, and some consume application characteristics (``Data source column''). Some agents (e.g., job performance estimators, HPC fault detection) need frequent retraining when infrastructure and job profile changes, while rule-based AI models require no training phase. 

Discrepancies in data sources, models' data sensitivity, and demand robust coordination and systematic MLOps workflows for potential integration. Recent studies are exploring best practices for integrating traditional software CI/CD pipelines with the machine learning and data lifecycle to develop comprehensive MLOps solutions \cite{mlops:2024}. In particular, methods that detect data distribution shifts by comparing historical data with incoming data, using similarity metrics, can effectively trigger the re-training of models \cite{mlops:2025}. However, the field of MLOps in the context of multi-heterogeneous agents is still relatively new.

They point to the growing necessity of standardization of MLOps practices and LLM-OS tailored for the large-scale nature of the HPC domain. Without such practices and systems, the promise of AI-enhanced HPC may be overshadowed by its operational overhead.

\textbf{Debugging and Monitoring}

Another critical challenge lies in the monitoring and reproducibility of AI agent interactions, particularly when those interactions span long timeframes or involve large, evolving datasets. As multiple agents exchange performance data, scheduling decisions, model predictions, and inferred states, the volume and variety of data traces quickly grow beyond what is human-auditable. Tracking who influenced what, when, and why becomes increasingly difficult.

This issue may become even more pronounced in scenarios involving interactions between two LLMs \cite{llm2llm:2024}, such as in the case of an LLM-OS coordinating with an LLM for code generation or other specialized tasks. The current LLM has a tendency for lengthy interaction, which significantly increases the log size.

This finding suggests that for AI models and their interactions, maintaining effective communication and logging requires substantial storage and computational resources. These demands, in turn, raise questions about the overall cost-benefit ratio of integrating such complex systems. The trade-off between the performance and adaptability gains from AI integration and the resource overhead involved in managing these interactions becomes a critical consideration.

\textbf{LLMS-OS Security}

The cybersecurity of LLMs and MLOps is an active area of research, with ongoing efforts addressing both offensive and defensive aspects. An LLM-OS, like any LLM-based service, typically exposes a chatbot graphical user interface to its users. As with any IT system, it is susceptible to traditional cybersecurity threats \cite{redhatllm:2024} such as unauthorized access, denial-of-service (DoS) attacks, privilege escalation, and man-in-the-middle attacks. Additionally, it is exposed to LLM-specific risks, including misuse for generating spam, scams, misinformation, violations of safety guardrails \cite{guard:2021, violguardrail:2025}, and potential data leakage \cite{infattack:2024}. In addition, MLOps in high-performance computing (HPC) environments introduces further concerns due to the experimental and rapidly evolving software stacks used in AI workflows  {   \cite{hpcsec:2024} }. 

The risks of deploying LLM-OS are further amplified in HPC systems. HPC systems typically manage vast datasets and demand significant computational power, resulting in high operational and environmental costs. Consequently, a security breach or system failure can have wide-reaching implications for institutions relying on these infrastructures. Consequently, the security LLM-OS of HPC systems deserves further evaluations and mitigations.

\section{Conclusion}
\label{sec:con}

This paper explores how various AI techniques can enhance High-Performance Computing (HPC). The analysis of 74 recent papers identified six key AI applications in HPC, including automatic job performance estimation, job scheduling, fault detection, performance optimization, surrogate models, and code/script automation through Natural Language Processing. Among these, job scheduling stands out as the most actively researched and in-demand area.

The key findings of this review are as follows:
\textbf{Key findings:}
\begin{itemize}
    \item \textbf{Performance estimation} is foundational, enabling informed scheduling, optimization, and even guiding code-generation agents.
    \item \textbf{Performance optimization} reduces runtime and energy while improving utilization across diverse workloads.
    \item \textbf{AI for job scheduling} is the most active topic, addressing distribution across increasingly heterogeneous resources. Approaches range from simulation-only research to production prototypes; while RL dominates, many systems combine supervised ML, recommendation, heuristics, and ILP.
    \item \textbf{AI-based surrogates} deliver substantial computational savings in simulations and DL, approximating HPC computations and, more recently, lower-level functions.
    \item \textbf{Fault detection and recovery} benefit from GNNs and time-series models (e.g., LSTM, Informer) that exploit spatial–temporal structure; most rely on minimal supervision but large production datasets.
    \item \textbf{Language Models for HPC automation} are advancing rapidly—LLMs enable test generation, code synthesis, and command-line automation. Models adapted to HPC contexts tend to outperform general-purpose ones on domain tasks.
\end{itemize}


The literature reveals interdependencies between several of these topics. For instance, AI-based performance estimators can support performance optimizers, inform job schedulers, and assist code/script generation tools. However, the integration of these individual AI applications into a unified and intelligent HPC framework has not yet been fully explored. This paper aims to contribute to that direction.

This analysis found that a particularly promising area of interest is the integration of Large Language Models (LLMs) into Operating Systems. This direction aims at more intuitive user interactions and allows the system to adapt dynamically to context, leading to more efficient resource utilization and potentially paving the way for the next generation of intelligent computing environments.


Despite its potential, integration raises several challenges:

{ 
\begin{itemize}
    \item  Lack of standardized metrics, datasets, and APIs hinders the integration and comparability of developed AI methods.
    \item AI components introduce new security concerns, particularly when integrated deeply into system-level infrastructure.
    \item Scaling MLOps workflows to accommodate frequent retraining and shifting data distributions.
    \item The live nature of data and models complicates debugging, reproducibility, and interpretability.
    \item Designing effective communication protocols among heterogeneous AI agents.
    \item Managing the computational and operational overhead introduced by logging, and coordination of large-scale AI interactions.
\end{itemize}
}

Given these challenges, the systematic benefits of such integration will need to be carefully evaluated in future research to determine whether the performance and adaptability gains justify the added complexity and risks.

{

HPC systems already enable major scientific breakthroughs in areas such as climate modeling, drug discovery, aerospace, automotive design, and energy optimization. By enhancing operational functions such as scheduling, anomaly detection, performance estimation and optimization, and surrogate modeling, AI can make these systems more efficient, reliable, and sustainable.
}

\clearpage

\bibliographystyle{elsarticle-num}

\bibliography{bib_aiworkload, bib_benchmark, bib_caffeine, bib_chara, bib_importance, bib_scheduling, bib_tuning, bib_rd, bib_ad, bib_surrogate, bib_review, bib_dataset, bib_intro, bib_llmos, bib_synergy, bib_survey}

\end{document}